\documentclass[a4paper,aps,prb,english,twocolumn,amsmath,showpacs,amssymb,nofootinbib]{revtex4-1}
\usepackage[pdftex]{color,graphicx,hyperref}
\pdfoutput=1
\usepackage[T1]{fontenc}
\usepackage[latin9]{inputenc}
\usepackage{babel}
\usepackage{amsmath}
\usepackage{amssymb}
\usepackage{graphicx}
\usepackage{bm}
\usepackage{amsbsy}

\hypersetup{
  colorlinks=true, 
  linkcolor=blue,
  citecolor=green,
  urlcolor=blue
}

\begin{document}

\title{Thermodynamics of a Potts-like model for a reconstructed zigzag
  edge in graphene nanoribbons}

\author{J. N. B. Rodrigues$^{1}$, P. A. D. Gon\c{c}alves$^{1,2}$,
  Jaime E. Santos$^{2,3,4}$ and A. H. Castro Neto$^{5,6}$}
\affiliation{$^1$ Centro de F\'{i}sica do Porto and Departamento de
  F\'{i}sica, Faculdade de Ci\^{e}ncias, Universidade do Porto, Rua do
  Campo Alegre 687, 4169-007 Porto, Portugal; $^2$ Centro de
  F\'{i}sica and Departamento de F\'{i}sica, Universidade do Minho,
  P-4710-057 Braga, Portugal; $^3$ Max Planck Institute for the
  Physics of Complex Systems, N\"othnitzer Str. 38, D-01187 Dresden,
  Germany; $^4$ Max Planck Institute for Chemical Physics of Solids,
  Noethnitzer Str. 40 D-01187 Dresden Germany; $^5$ Graphene Research
  Centre and Physics Department, National University of Singapore, 2
  Science Drive 3, Singapore 117542; $^6$ Department of Physics,
  Boston University, 590 Commonwealth Avenue Boston MA 02215 USA}

\email{joaonbrod@gmail.com}


\date{\today}

\begin{abstract}
  We construct a three-color Potts-like model for the graphene zigzag
  edge reconstructed with Stone-Wales carbon rings, in order to study
  its thermal equilibrium properties. We consider two cases which have
  different ground-states: the edge with non-passivated dangling
  carbon bonds and the edge fully passivated with hydrogen. We study
  the concentration of defects perturbing the ground-state
  configuration as a function of the temperature. The defect
  concentration is found to be exponentially dependent on the
  effective parameters that describe the model at all
  temperatures. Moreover, we analytically compute the domain size
  distribution of the defective domains and conclude that it does not
  have fat-tails. In an appendix, we show how the exchange parameters
  of the model can be estimated using density functional theory results. Such equilibrium
  mechanisms place a lower bound on the concentration of defects in
  zigzag edges, since the formation of such defects is due to
  non-equilibrium kinetic mechanisms.
\end{abstract}

\maketitle

\section{Introduction}

The mechanical exfoliation of graphene,\cite{Novoselov_Science:2004} a
one-atom thick sheet of carbon atoms arranged in a honeycomb
structure, followed by subsequent experiments,
\cite{Novoselov_PNAS:2005, Novoselov_Nature:2005, Zhang_Nature:2005}
has given rise to a profusion of studies, both experimental and
theoretical (see Refs. \onlinecite{Neto_RMP:2009, Peres_RMP:2010,
  Jia_Nanoscale:2011, Molitor_JPCM:2011, Sarma_RMP:2011} for a list of
references) regarding the physical properties of this remarkable
material.  Such an interest has not subsided to the present date,
quite on the contrary.

A simple nearest-neighbor tight-binding approximation of the
electronic Hamiltonian in graphene reveals that the honeycomb lattice
structure leads to a dispersion relation that is linear around two
specific points of the Brillouin zone, the Dirac points. Since the
Fermi level of pristine graphene lies at these points, its
quasi-particles behave, in a continuum approximation, as mass-less
relativistic fermions with a speed of light equal to the
Fermi-velocity ($v_{F} \approx 10^{6} m
s^{-1}$).\cite{Wallace_PR:1947, Neto_RMP:2009}

Typically, graphene exhibits high crystal quality, large transparency,
being highly conductive and very strong yet
flexible.\cite{Neto_RMP:2009, Peres_RMP:2010} All these
characteristics place graphene as a good candidate to be used in a
variety of technological applications, such as in solar cell
technology,\cite{Wang_NL:2008} in liquid crystal
devices,\cite{Blake_NL:2008} in single molecule
sensors,\cite{Schedin_NatM:2007} in the fabrication of nano-sized
prototype transistors,\cite{Wang_PRL:2008} among many
others. Understanding the transport properties of graphene is thus an
essential research program towards its application to future
nanoscopic devices.

\begin{figure}[htp!]
  \centering
  \includegraphics[width=0.98\columnwidth]{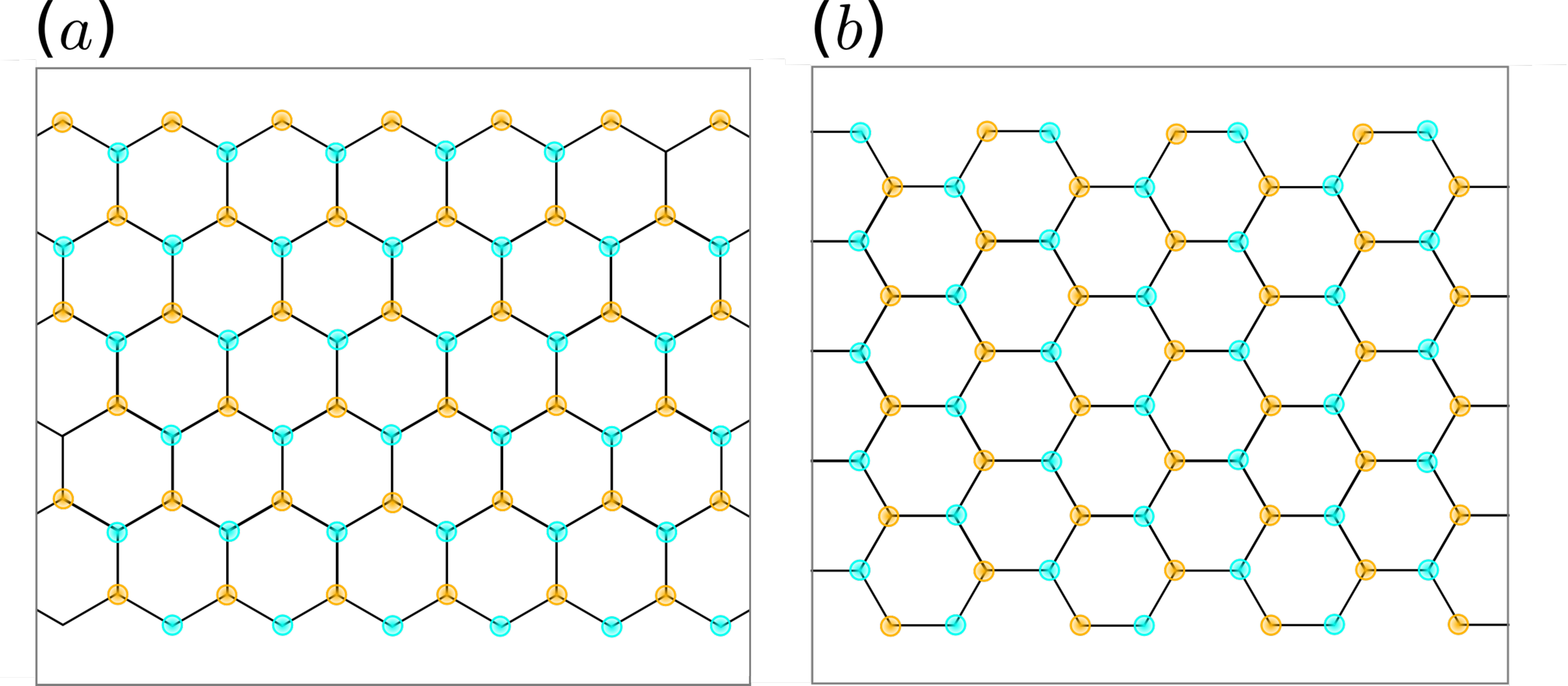}
  \caption{(Color online) (a) Scheme of zigzag ribbon. (b) Scheme of
    an armchair ribbon.}
  \label{fig:GNRs}
\end{figure}

Several of these sought nanoscopic devices will certainly be based on
the use of graphene ribbons and graphene quantum dots. In particular,
graphene ribbons are usually classified as zigzag or armchair,
depending on their edge configuration (see Fig. \ref{fig:GNRs}). It is
already well established that the electronic properties of these
nanostructures are strongly affected by their edge configuration. A
nearest-neighbor tight-binding approach leads to the conclusion that
zigzag ribbons are metallic regardless of their width, while
armchair ribbons can be either semiconducting or metallic, depending
on their width. In addition, zigzag ribbons present edge localized
states around the Fermi energy.\cite{Fujita_JPSJ:1996,
  Nakada_PRB:1996, Wakabayashi_PRB:1999} However, {\it ab-initio}
calculations predict that graphene ribbons are always
semiconducting.\cite{Barone_NL:2006} Experimental results show that
the ribbons' energy gaps increase with decreasing ribbon
width.\cite{Son_Nature:2006}

It has been shown\cite{Lewenkopf_PRB:2008, Mucciolo_PRB:2009} that
edge disorder modifies substantially the electronic properties of
nanostructures based on graphene and is responsible for most of the
transport properties in these systems.  The wave-functions associated
with zigzag edges and vacancies\cite{Pereira_PRL:2006} decay very
slowly with the distance from the edge because of the absence of a gap
in the graphene spectrum. This slow decay leads to strong quantum
interference effects that are responsible for destructive quantum
interference, Anderson localization, and Coulomb blockade effects in
the electrical conductance of graphene-based
devices.\cite{Sols_PRL:2007} While most of edge disorder in graphene
nanostructures is produced by the method of cutting graphene by hot
plasma,\cite{Han_PRL:2007, Ponomarenko_Science:2008, Han_PRL:2010} it
also shows conspicuously in chemically driven methods which can be
classified as quasi-equilibrium.\cite{Xie_JACS:2011} It thus follows
that equilibrium mechanisms, such as those described in this work,
only place a lower bound on the amount of disorder that can exist in
these nanostructures and hence an upper limit in the value of the
conductance that can be obtained in these devices.

It is thus of extreme importance for graphene electronics development
to understand the effect of edge disorder in the transport properties
of graphene ribbons and quantum dots.

One example of disorder appearing at graphene edges are sets of five
and seven sided rings of carbon atoms, commonly named in the literature as
Stone-Wales (SW) defects.\cite{Stone_CPL:1986} We emphasize that, for the sake 
of clarity, from now on, we are going to name these structures as {\it Stone-Wales 
carbon rings}, limiting the use of the word {\it defects} to the 
context of (thermal) disorder. 

The SW carbon rings have been observed and found to be meta-stable in the bulk 
of graphene sheets.\cite{Meyer_NL:2008} Moreover, and regarding graphene's 
high temperature behavior, it was found that the formation of Stone-Wales
carbon rings is the first step in the process of graphene's melting
($T_{melting} \approx 4900 \mbox{K}$).\cite{Zakharchenko_JPCM:2011}
Besides, it has been shown through {\it ab-initio} calculations that
whenever SW carbon rings are present in non-passivated graphene
nanoribbons, the ribbons energy decreases as the carbon ring approaches 
the edge of the ribbon.\cite{Oeiras_PRB:2009} In addition, further 
{\it ab-initio} calculations have also shown that the formation of SW
carbon rings at the edges of both armchair and zigzag nanoribbons,
stabilize them, both energetically and mechanically.
\cite{Koskinen_PRL:2008, Huang_PRL:2009, Bhowmick_PRB:2010} In
particular, in the absence of hydrogen passivation, the zigzag edge is only a
meta-stable state, the state where the edge is fully reconstructed with SW 
carbon rings being the ground-state of the system.\cite{Huang_PRL:2009} 
Moreover, such total reconstruction of the zigzag edge gives rise to
the appearance of a new kind of edge state.\cite{Rodrigues_PRB:2011,Ostaay_PRB:2011}
However, if the zigzag edges are hydrogen passivated, it is the perfect 
zigzag edge that has the lower energy.  The reconstruction of the zigzag 
edge by SW carbon rings acts as a mechanism that self-passivates the edge.
\cite{Koskinen_PRL:2008} Density functional theory and molecular dynamics 
calculations corroborate these results, pointing to an energy barrier associated
with the edge reconstruction of about $0.4$-$0.9 e\mbox{V}$ per edge
unit cell.\cite{Koskinen_PRL:2008, Lee_PRB:2010, Li_PRB:2010,Kroes_PRB:2011} 
 
\begin{figure}[htp!]
  \centering
  \includegraphics[width=0.98\columnwidth]{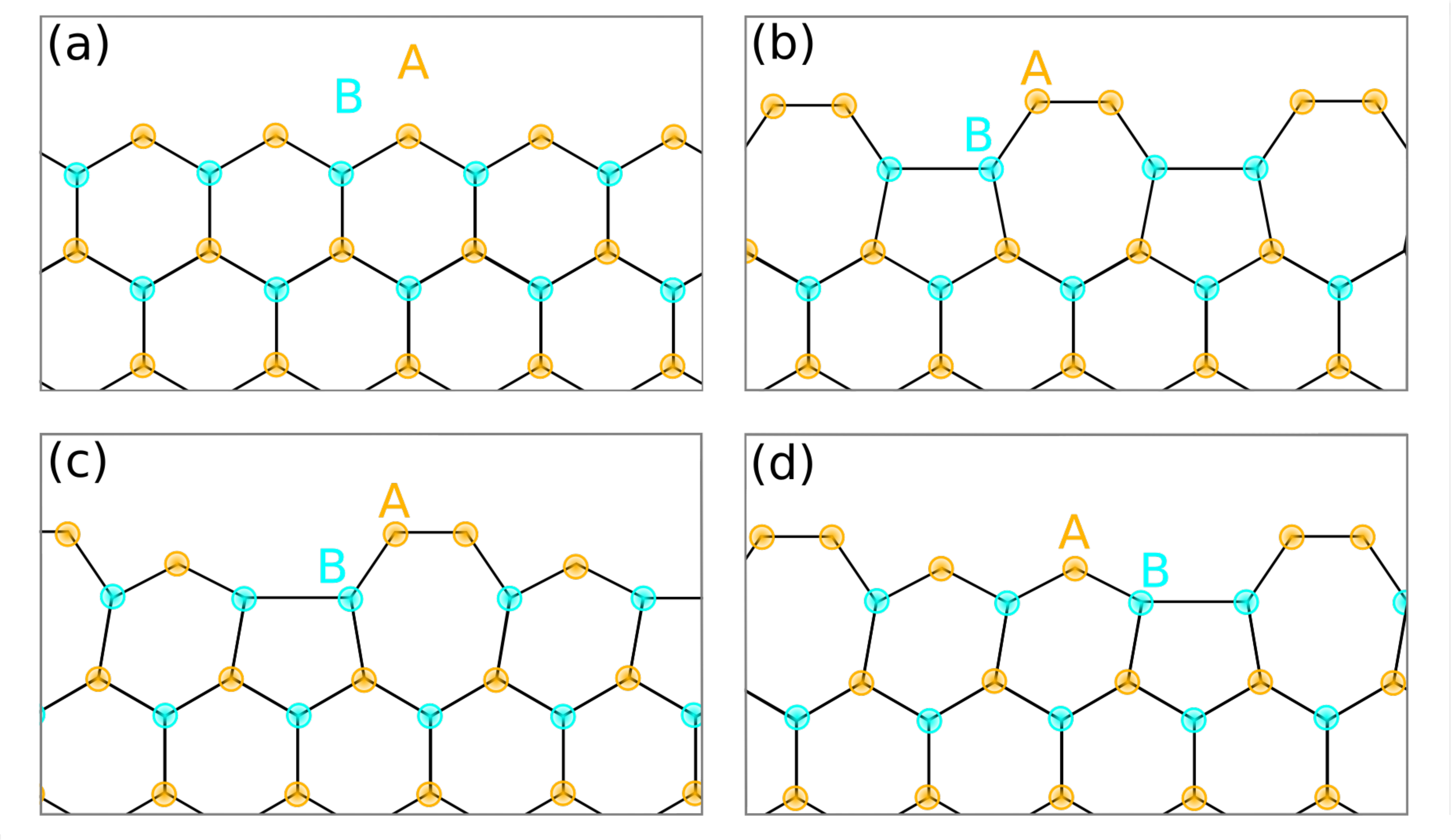}
  \caption{(Color online) (a) Scheme of clean zigzag edge, usually named
    $zz$. (b) Scheme of a totally reconstructed (with SW carbon rings)
    zigzag ribbon, usually named $zz(57)$. In (c) and (d), we
    present schemes of a partially reconstructed zigzag ribbon,
    respectively, $zz(576)$ and $zz(5766)$.}
  \label{fig:reczag_family}
\end{figure}

These types of zigzag edge reconstructions (see
Fig. \ref{fig:reczag_family}), are claimed to be stable only at very
low hydrogen pressure (well below ambient conditions) and very low
temperatures.\cite{Wassmann_PRL:2008} However, reconstructions of the
zigzag (as well as armchair) edges have been recently observed with
high-resolution transmission electron microscopy (TEM),\cite{Girit_Science:2009, Chuvilin_NJP:2009,
  Koskinen_PRB:2009} albeit under rather extreme conditions, namely,
the graphene flake is bombarded with high-energy electrons ($80\,
\mbox{k}e\mbox{V}$) that remove $C$ atoms from the sheet. The recent
work of Suenaga {\it et al.},\cite{Suenaga_Nature:2010} on single-atom
spectroscopy using low-voltage scanning TEM (STEM), provides a {\it non-destructive}
method of identifying the edge configuration of graphene ribbons, as
does the work of Warner {\it et al.} on the observation of real-time
dynamics of dislocations using high-resolution
TEM.\cite{Warner_Science:2012} Moreover, refinements in other
techniques, such as Raman spectra of the edges,\cite{Malola_EPJD:2009}
scanning tunneling microscopy (STM) images of the edges, \cite{Koskinen_PRL:2008} or coherent electron
focusing\cite{Rakyta_PRB:2010} may help in identifying edge reconstructions.

In this work, we construct a one-dimensional three-color (or three-state) Potts-like
model so as to describe the reconstruction of the zigzag edge with SW
carbon rings, due to thermal fluctuations. From such a model, we will
extract thermodynamic properties of the zigzag edge, both in the
absence and in the presence of hydrogen passivation. These properties
include the number of pentagons and heptagons at an edge, 
the concentration of domains formed by pentagon-heptagon 
pairs within larger groups of hexagons or the concentration of domains 
formed by hexagons within larger groups of pentagon-heptagon 
pairs (depending on the state of passivation of the edge), and the size-distribution 
of both kinds of minority domains at the edge. These quantities are then used to 
characterize the degree of disorder of the zigzag edge due to thermal fluctuations.

The simple model presented in this work indicates that if the nanoribbon edge behaves as a 
one-dimensional system with short-range interactions, it will present a finite concentration of defects 
at any finite temperature. Moreover, the average length of such defects is finite at any non-zero
temperature. This is in sharp contrast with the case of the one-dimensional ferromagnetic Ising model, 
where domains of minority spins, which can be viewed as the defects perturbing the ground-state
configuration, have an average length that diverges at infinitesimally small temperatures, 
destroying the ferromagnetic order. We believe that such distinct behavior is due to the lack of a 
full $\mathbf{Z}_{3}$ symmetry in the three-color Potts-like model. Furthermore,
we are able to compute analytically the distribution of lengths of defective domains (DSD), as well as 
the concentration of such domains. We show that the DSD has no fat-tails. 

From an experimental point of view, the concentration of defective domains is the most relevant quantity that 
one can compute, since it is found to be exponentially dependent on the values of the effective exchange 
parameters of the model. Hence, the measurement of the concentration of defects would allow for a sensitive 
determination of these parameters. Since such measurements have not yet been performed, we have estimated these 
parameters using density functional theory (DFT) calculations. Depending on the actual value of the effective parameters, the 
concentration of defects can become quite large at room temperatures, and may thus have a significant effect
on the conductivity of the zigzag ribbon.

The structure of this paper is as follows: Section \ref{sec:PottsModel} will be devoted 
to the presentation of the three-color Potts-like model that describes the thermodynamics 
of the edge with SW carbon rings and of the results extracted from it. We will first present 
a descriptive outline of the model. In sub-Section \ref{sec:EdgeThermo}, we will compute the
thermodynamic quantities characterizing the edge, using a transfer
matrix formulation of the Potts-like model and we will analyze their
dependence both on the temperature and on the exchange
parameters. Finally, in Section \ref{secCon}, we will present our
conclusions. We leave to the appendices the computation of the exchange parameters from {\it ab-initio}
results (Appendix \ref{sec:ExchangeParams}), the calculation of correlation functions in the three-color
Potts-like model, using the transfer matrix formalism (Appendix \ref{appA}), and the
explicit computation of the size-distribution of domains of
polarized and unpolarized spins (Appendix \ref{app:DSD}).

\section{Potts-like model of the zigzag edge}

\label{sec:PottsModel}

\begin{figure}[htp!]
  \centering
  \includegraphics[width=0.98\columnwidth]{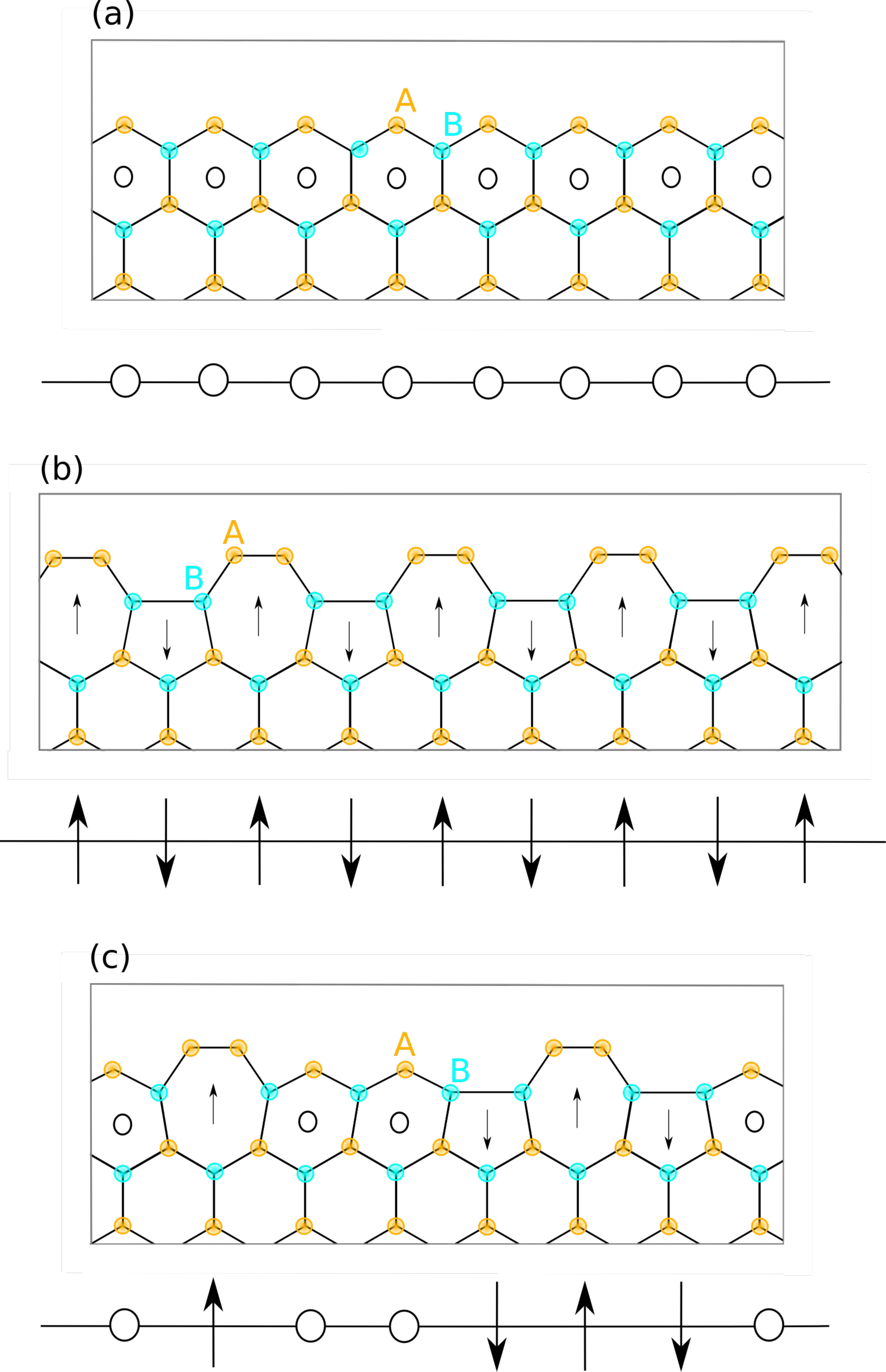}
  \caption{(Color online) Scheme of the Potts-like model (three-color) for
    the zigzag edge with SW carbon rings. In (a), a clean zigzag
    edge, $zz$, is shown.  In (b), a zigzag edge which is totally
    reconstructed with SW carbon rings, $zz(57)$, is presented. In
    (c), a zigzag edge with an arbitrary reconstruction is shown.}
  \label{fig:PottsModel_EdgeSWDs}
\end{figure}

We will study graphene zigzag edges with SW carbon rings, employing a
one-dimensional three-color Potts-like model, where each color is
assigned to a different polygon of the edge (hexagons, heptagons and
pentagons).  We label each edge unit cell, i. e. each polygon at the 
edge, by the integer variable $i = 0, 1, \cdots, 2N$, with the
state of such a cell being described by the ternary variable
$\sigma_{i} = 0, +1, -1$, according to whether the polygon forming that
cell in the reconstructed edge is an hexagon, heptagon or pentagon.
We consider a nearest-neighbor coupling between adjacent cells only
(the validity of this assumption will be justified in Appendix
\ref{sec:ExchangeParamsResults}), which leaves us with $9$ possible
values for the couplings $J_{\sigma_{i} \sigma_{i+1}}$, depending on
the neighboring states. We take as reference state with zero energy
the perfect zigzag edge, thus $J_{00}=0$. Taking into account the
experimental observations, we will exclude from the model states where
two pentagons or heptagons sit at neighboring sites, i.e. pairings of
heptagons or pentagons are forbidden and one has
$J_{++}=J_{--}=\infty$.  Invariance under inversion implies that the
order in a pentagon-heptagon, pentagon-hexagon or heptagon-hexagon
pair is irrelevant, and thus $J_{-+}=J_{+-}$, $J_{0+}=J_{+0}$ and
$J_{0-}=J_{-0}$.  Moreover, since heptagons and pentagons are created
in pairs through the transference of $C$ atoms between neighboring
sites, we will assume that the probability of creation of a pentagon
or an heptagon is the same, which implies that $J_{0-}=J_{0+}$. Hence,
the $9$ initial possible values of the couplings are reduced to two
free parameters: $J_{0+}=\gamma>0$, which reflects the fact that the
formation of defects costs energy and $J_{+-}=\delta$, which may be
negative or positive depending on whether the totally reconstructed
edge has lower or higher energy than the pristine zigzag edge (i.e.,
depending on the state of passivation of the edges, as discussed
above). Finally, since $C$ atoms are conserved, one should, strictly
speaking, consider a model with as many heptagons as pentagons,
i.e. one should work in a subspace of the state-space having the
overall magnetization $M=\sum_{i=0}^{2N}\,\sigma_i=0$. Such a
constraint can be written in terms of an imaginary applied magnetic
field over which one has to integrate, once the eigenvalues of the
transfer matrix of the Potts-like model have been computed. In such a
case, the eigenvalues can no longer be simply determined. We will
therefore relax this constraint and we will only implement it on
average, as $\langle\, M \,\rangle=0$ in one-dimension (1D). Note, moreover, that
although some of the edge observation
techniques\cite{Girit_Science:2009, Chuvilin_NJP:2009,
  Koskinen_PRB:2009} are highly energetic and cause the ejection of
$C$ atoms from the edges, the system cannot be considered to be in
thermodynamic equilibrium when such ejection occurs and the model
introduced below is therefore not applicable.\cite{Girit_Science:2009}
It may however be applicable after a characteristic relaxation time,
such that the thermodynamics of the edge would be described in terms
of an effective temperature, dependent on the energy deposited by the
electron beam and the heat conduction process in graphene. One would
expect the number of (remaining) $C$ atoms in the edge to be conserved
in this late-time regime. In Fig. \ref{fig:PottsModel_EdgeSWDs}, we
present a cartoon of three possible configurations of the edges and
how they translate into configurations of the three-color Potts-like
model.

\subsection{Edge thermodynamics}

\label{sec:EdgeThermo}

In the previous paragraphs, we have shown how to map the different
configurations of a reconstructed edge of a graphene zigzag ribbon to
those of a three-color nearest-neighbor Potts-like model. We now wish to 
use such a model to compute useful 
quantities relating to the thermodynamics of the edges. As is usual in
one-dimensional models with nearest neighbor interactions, the
speediest way to compute thermodynamic properties, including spin-spin
correlation functions, is to express these quantities in terms of a
transfer matrix.\cite{Huang_Book:1987} In the case of the model
presented above, the transfer matrix reads:
\begin{equation}
  \label{eq1}
  \mathbf{T}=\left( \begin{array}{ccc}
      0 & e^{-\beta \gamma}&e^{-\beta \delta}  \\
      e^{-\beta \gamma}& 1&e^{-\beta \gamma}\\
      e^{-\beta \delta}&e^{-\beta \gamma} & 0 
    \end{array} \right)\enspace\,,
\end{equation}
where $\beta=1/k_B T$ and $\gamma$ and $\delta$ are the parameters
introduced above and computed in Appendix \ref{sec:ExchangeParams}. The model, as
defined by Eq. (\ref{eq1}), represents a limiting case of the
Blume-Emery-Griffiths model in one dimension.
\cite{Rys_HPA:1969,Hintermann_HPA:1969,Blume_PRA:1971} One eigenvalue
of the transfer matrix, $\lambda_0=-e^{-\beta\delta}$, can be readily
identified, after which the other two are also easily computed from
the quadratic equation that is obtained from the application of, e.g.,
Ruffini's rule to the cubic secular equation. These two eigenvalues are
given by $\lambda_{\pm}=\frac{1}{2}\left[1+e^{-\beta\delta}\pm\sqrt{
    (1-e^{-\beta\delta})^2+8\,e^{-2\beta\gamma}}\right]$. At all
temperatures above zero, $\lambda_+$ is the largest eigenvalue. At
$T=0$ and if $\delta>0$, this is also the case, however if $\delta\leq
0$ the largest eigenvalue may be doubly or thrice degenerate, which
reflects the degeneracy of the ground-state of the system (see
Appendix \ref{appA}).

The free energy of the system is given by
$F=-k_BT\ln(\mbox{Tr}\,\mathbf{T}^{2N})$ from which we have that in
the thermodynamic limit $N\rightarrow\infty$ the free energy per site
is simply proportional to the logarithm of the largest eigenvalue,
i.e.
\begin{equation}
  \label{eq2}
  f=-k_BT\ln\left\{\frac{1}{2}\left[1+e^{-\beta\delta}+\sqrt{
        (1-e^{-\beta\delta})^2+8\,e^{-2\beta\gamma}}\right]\right\}\,.
\end{equation}

We are primarily interested in the disorder caused either to a clean zigzag edge (ground-state 
of the passivated edge) or to a totally reconstructed zigzag edge (ground-state of the non-passivated edge) 
through the effect of temperature, which leads these configurations [as depicted in Fig. 
\ref{fig:PottsModel_EdgeSWDs}(a) and (b)] to evolve into Fig. \ref{fig:PottsModel_EdgeSWDs}(c). For a 
totally passivated edge, a measure of such disorder can be obtained by counting the number 
of domains of polarized spins (heptagons and pentagons) that exist between sites with $0$-spin 
(hexagons), the converse being valid for the non-passivated edge. As an example, in
Fig. \ref{fig:PottsModel_EdgeSWDs}(c) one has two domains of
polarized spins. In order to be able to count them, consider the
contribution of a domain, both to $\sum_i \sigma_i^2$, which measures
the number of heptagons or pentagons in the system, and to $\sum_i
\delta_{\sigma_i\sigma_{i+1},-1}$, which measures the number of
heptagon-pentagon links (see Table \ref{table:1}).
\begin{table}[t]
  \begin{center}
    \begin{tabular}{|c|c|c|c|}
      \hline
      Configuration & $\sum_i' \sigma_i^2$ &$\sum_i' \delta_{\sigma_i\sigma_{i+1},-1}$ & Difference\\ \hline
      $0\,+\,0$&1&0&1 \\ \hline
      $0\,-\,0$&1&0&1 \\ \hline 
      $0\,+\,-\,0$&2&1&1 \\ \hline
      $0\,-\,+\,0$&2&1&1 \\ \hline
      $0\,+\,-\,+\,0$&3&2&1 \\ \hline
      $0\,-\,+\,-\,0$&3&2&1 \\ \hline
    \end{tabular}
  \end{center}
  \caption{Contribution of a $+-$ domain for the different observables.}
  \label{table:1}
\end{table}
Since each domain contributes exactly 1 to the difference between
these two quantities, one sees that the number of domains is given by
the difference of these two operators. Note however, that whenever the
spin chain has no $0$-spins, the difference between these two
operators gives $0$. As a consequence, the correct expression for the
average domain concentration of $\pm$-spin domains, $\langle n_{d \pm}
\rangle = \langle N_{d \pm} \rangle /2 N$, is given by
\begin{equation}
  \label{eq3}
  \langle n_{d \pm} \rangle = \frac{1}{2N}\, \bigg[ \sum_i\, \Big(\langle\,\sigma_i^2\,
  \rangle-\langle\,\delta_{\sigma_i\sigma_{i+1},-1}\,\rangle \Big) + \Big\langle \prod_{i} \sigma_i^2 \Big\rangle
  \bigg]\,.
\end{equation}

If one uses periodic boundary conditions (PBCs) and the spin chain is
not uniformly polarized (either all the sites with spin $0$, or all
the sites with alternating polarized spins $+-+-+-$), the number of domains of
$\pm$-spins is always equal to the number of domains of $0$-spins. In
such a case, we have $N_{d \pm} = N_{d 0} \equiv N_{d}$. As a
consequence, we can express $\langle n_{d 0} \rangle$ in terms of
$\langle n_{d \pm} \rangle$, just by considering the following sum
over all spin configurations,
\begin{eqnarray}
  \langle n_{d 0} \rangle &=& \langle n_{d \pm} \rangle - \frac{1}{2 N} \bigg[ \Big\langle \prod_{i} 
  \sigma_i^2 \Big\rangle - \Big\langle \prod_{i} \big( 1 - \sigma_i^2 \big) \Big\rangle \bigg] .
  \label{eq:Relation_nd0_ndpm}
\end{eqnarray}
Note that in the thermodynamic limit, $2 N \to \infty$, the thermal
averages of the products can be neglected, resulting in $\langle n_{d
  0} \rangle \approx \langle n_{d \pm} \rangle$.

One can separately compute the correlation functions
$\langle\,\sigma_i^2\, \rangle$ and
$\langle\,\delta_{\sigma_i\sigma_{i+1},-1}\,\rangle$, as is done in
Appendix \ref{appA}. However, it is simpler to consider instead
generating fields in the partition sum that are coupled to
$\sum_i\,\sigma_i^2$ and to
$\sum_i\,\delta_{\sigma_i\sigma_{i+1},-1}$.  One then concludes, using
the transfer matrix formalism, that the concentration of polarized sites,
$\langle n_{pol} \rangle = \frac{\langle N_{pol} \rangle}{2N} =
\frac{1}{2N}\,\sum_i\langle\, \sigma_i^2\,\rangle$, is given, in the
thermodynamic limit, by
\begin{equation}
  \langle n_{pol} \rangle =\frac{1}{2}\frac{\partial f}{\partial \gamma}+\frac{\partial f}{\partial \delta}\,.
  \label{eq3a}
\end{equation}
The concentration of unpolarized sites is simply obtained from $\langle
n_{unp} \rangle = 1 - \langle n_{pol} \rangle$.  Moreover, the concentration
of links between polarized sites, defined as, $\langle n_{+-} \rangle
=\frac{\langle
  N_{+-}\rangle}{2N}=\frac{1}{2N}\,\sum_i\langle\,\delta_{\sigma_i\sigma_{i+1},-1}\,\rangle$,
is given, in the thermodynamic limit, by
\begin{equation}
  \langle n_{+-} \rangle =\frac{\partial f}{\partial \delta}\,.
  \label{eq3b}
\end{equation} 
Similarly, the concentration of links between unpolarized sites, is simply
obtained from $\langle n_{00} \rangle = 1- \langle n_{+-} \rangle -
\langle n_{\pm 0} \rangle$, where $n_{\pm 0}$ stands for the concentration
of links between polarized and unpolarized sites. Note that in the
thermodynamic limit, $\langle n_{\pm 0} \rangle \approx 2 \langle n_{d
  0} \rangle \approx 2 \langle n_{d \pm} \rangle$.  
  
If one substitutes in equation \eqref{eq3} the 
expression for $\langle n_{pol} \rangle$ and for $\langle n_{+-} \rangle$, as
given by Eqs. (\ref{eq3a}) and (\ref{eq3b}), one can write for 
$\langle n_{d \pm} \rangle$, in the thermodynamic limit, the result
\begin{equation}
  \label{eq4}
  \langle n_{d \pm} \rangle = \frac{1}{2}\frac{\partial f}{\partial \gamma}\,.
\end{equation}

\begin{figure}[htp!]
  \centering
  \includegraphics[width=0.98\columnwidth]{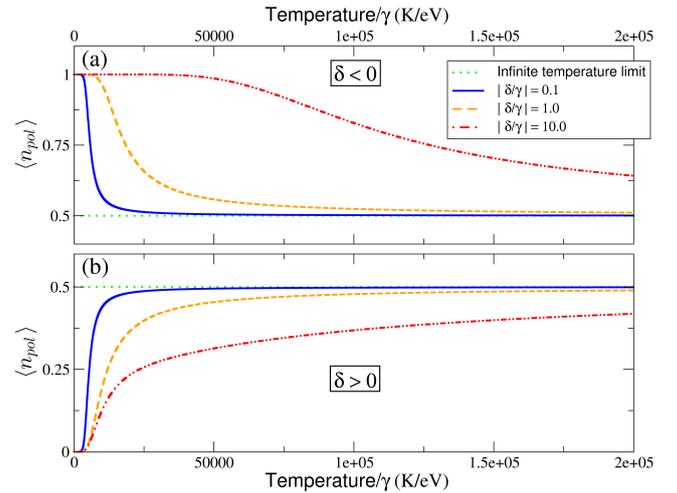}
  \caption{(Color online) Plot of the concentration of polarized
    spins, $\langle n_{pol} \rangle$, either for a non-passivated edge
    (a) and for a hydrogen-passivated one (b) as a function of
    $T/\gamma$, for three different values of the ratio
    $\delta/\gamma$. The full dark blue lines stand for $|
    \delta/\gamma | = 0.1$; the dashed orange lines stand for $|
    \delta/\gamma | = 1.$; the dashed-dotted red lines stand for $|
    \delta/\gamma | = 10.$. Note that the $\delta/\gamma$ ratio is
    negative for the non-passivated case (because $\delta < 0$) and
    positive for the passivated case (because then $\delta > 0$). The
    green dotted flat line represents the infinite temperature limit
    for both cases, which is $1 / 2$. Since we are using PBC, $\langle
    n_{unp} \rangle = 1 - \langle n_{pol} \rangle$.}
  \label{fig:npol}
\end{figure}

Substituting Eq. (\ref{eq2}) in Eq. (\ref{eq3a}), we obtain for
$\langle n_{pol} \rangle$
\begin{eqnarray}
  \langle n_{pol} \rangle &=& \frac{ 4 e^{-2 \beta \gamma} + e^{- \beta \delta} 
    (-1 + e^{-\beta \delta} + \theta) }{ (1 + e^{-\beta\delta} + \theta) 
    \theta } , \label{eq:polDensity}
\end{eqnarray}
where $\theta = \sqrt{(1 - e^{-\beta\delta})^2 + 8
  e^{-2\beta\gamma}}$. A plot of this quantity as a function of
$T/\gamma$, for selected values of the ratio $\delta/\gamma$, is shown in
Fig. \ref{fig:npol}.

\begin{figure}[htp!]
  \centering
  \includegraphics[width=0.98\columnwidth]{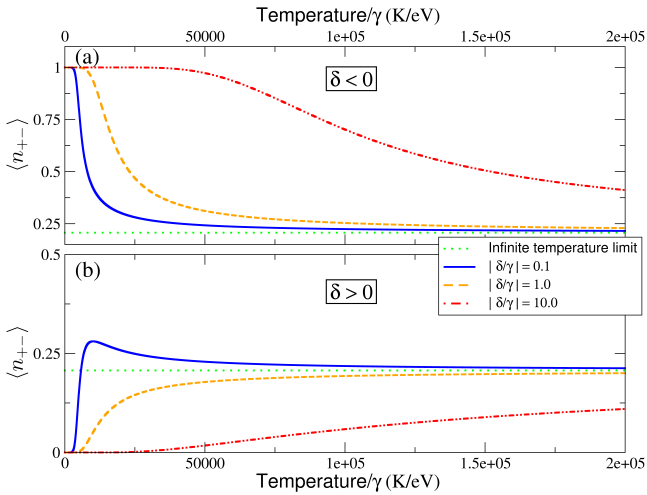}
  \caption{(Color online) Plot of the concentration of links between
    polarized spins, $\langle n_{+-} \rangle$, either for a
    non-passivated edge (a) and for a hydrogen-passivated one (b) as a
    function of $T/\gamma$, for three different values of the ratio
    $\delta/\gamma$. The full dark blue lines stand for $|
    \delta/\gamma | = 0.1$; the dashed orange lines stand for $|
    \delta/\gamma | = 1.$; the dashed-dotted red lines stand for $|
    \delta/\gamma | = 10.$.  Note that the $\delta/\gamma$ ratio is
    negative for the non-passivated case (because $\delta < 0$) and
    positive for the passivated case (because then $\delta > 0$). The
    green dotted flat line represents the infinite temperature limit
    for both cases, which is $1 / (2 + 2 \sqrt{2})$. Since we are
    using PBC, $\langle n_{00} \rangle = 1 - \langle n_{+-} \rangle -
    \langle n_{\pm 0} \rangle$.}
  \label{fig:npmb}
\end{figure}

Substituting Eq. (\ref{eq2}) in Eq. (\ref{eq3b}), we obtain for
$\langle n_{+-} \rangle$
\begin{eqnarray}
  \langle n_{+-} \rangle &=& \frac{ e^{- \beta \delta} (-1 + e^{-\beta \delta} + \theta) }
  { (1 + e^{-\beta\delta} + \theta) \theta } \,. \label{eq:pmbDensity}
\end{eqnarray}
A plot of Eq. \eqref{eq:pmbDensity} as a function of $T/\gamma$, for selected
values of the ratio $\delta/\gamma$, is shown in Fig. \ref{fig:npmb}.

\begin{figure}[htp!]
  \centering
  \includegraphics[width=0.98\columnwidth]{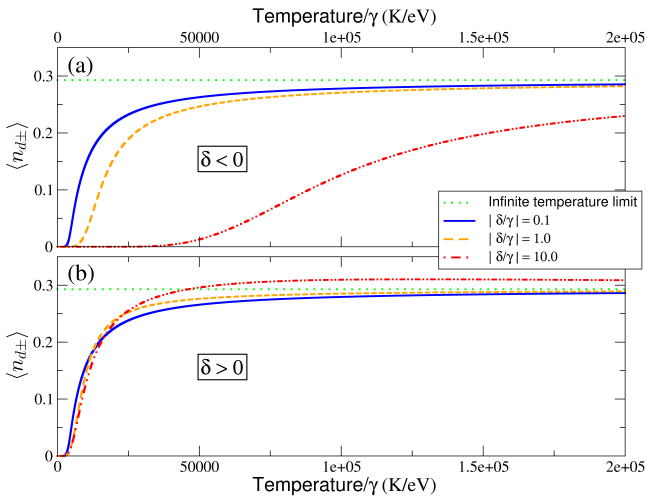}
  \caption{(Color online) Plot of the concentration of domains of
    polarized spins, $\langle n_{d \pm} \rangle$, for a non-passivated
    edge (a) and a hydrogen-passivated one (b) as a function of
    $T/\gamma$, for three different values of the ratio
    $\delta/\gamma$. The full dark blue lines stand for $|
    \delta/\gamma | = 0.1$; the dashed orange lines stand for $|
    \delta/\gamma | = 1.$; the dashed-dotted red lines stand for $|
    \delta/\gamma | = 10.$.  Note that the $\delta/\gamma$ ratio is
    negative for the non-passivated case (because $\delta < 0$) and
    positive for the passivated case (because $\delta > 0$ in such a
    case). The green dotted flat line represents the infinite
    temperature limit for both cases, which is $1 / (2 +
    \sqrt{2})$. Since we are using PBCs, in the thermodynamic limit,
    $\langle n_{d 0} \rangle \approx \langle n_{d \pm} \rangle$.}
  \label{fig:nd}
\end{figure}

Finally, substituting Eq. (\ref{eq2}) in Eq. (\ref{eq4}), we obtain
for $\langle n_{d \pm} \rangle$
\begin{equation}
  \label{eq5}
  \langle n_{d \pm} \rangle = \frac{4 e^{-2\beta \gamma}}{(1 + e^{- \beta \delta} + \theta) \theta} ,
\end{equation}
which gives the concentration of domains of polarized spins as a function of the 
temperature and of the coupling parameters. A plot of $\langle n_{d \pm} \rangle$
as a function of $T/\gamma$, is shown for selected values of the ratio
$\delta/\gamma$ in Fig. \ref{fig:nd}.

In both the non-passivated case and the hydrogen-passivated one, the
concentration of domains of polarized sites, $n_{d}$, is very small at low
temperatures (Fig. \ref{fig:nd}). However, from Figs. \ref{fig:npol}
and \ref{fig:npmb}, we conclude that these two situations are
substantially different. In the former, at low temperature, we have a
small number of very large polarized domains, with very few $0$-spins
between them.  In contrast, in the latter case, at low temperature, we
have a low number of very small polarized domains, with large domains
of $0$-spins between them. This is merely a manifestation of the fact 
that the two cases have different ground-states.

\begin{figure}[htp!]
  \centering
  \includegraphics[width=0.98\columnwidth]{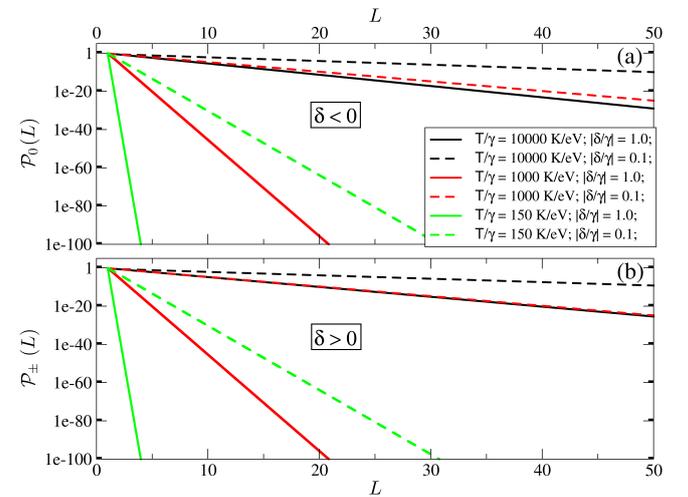}
  \caption{(Color online) Plot of the DSD, $\mathcal{P}(L)$, for
    several temperatures and several values of the ratio
    $\delta/\gamma$. The DSD is plotted with a logarithmic scale in
    the $y$-axis. (a) DSD of unpolarized spins in a
    non-passivated edge ($\gamma > 0$ and $\delta < 0$). (b) DSD 
    of polarized spins in a hydrogen-passivated edge ($\gamma
    > 0$ and $\delta > 0$). The black, red, and light green curves
    stand, respectively, for $T/\gamma=10000$ K/eV, $T/\gamma=1000$
    K/eV and $T/\gamma=150$ K/eV. The full and dashed lines, stand,
    respectively, for $\vert \delta/\gamma \vert = 1.0$ and $\vert
    \delta/\gamma \vert = 0.1$.}
  \label{fig:DSDgen}
\end{figure}

The DSD of polarized spins,
$\mathcal{P}_{\pm} (L) = \langle N_{d_L}^{\pm} / N_{d \pm} \rangle$,
where $L$ is the length of the domain and $N_{d_L}^{\pm}$ is the
number of $+-$ domains with size equal to $L$, can be computed exactly
using the transfer matrix formalism for the three-states Potts-like model
developed above. The detailed (and rather lengthy) calculation is
presented in Appendix \ref{app:DSD}. We obtain, in the thermodynamic
limit, the result
\begin{eqnarray}
  \mathcal{P}_{\pm} (L) &=& \frac{\lambda_{+} e^{\beta \delta} - 1}{\big( \lambda_{+} e^{\beta \delta} \big)^{L}}\,. 
  \label{eq:DSDpolTh}
\end{eqnarray}
Likewise, we have also computed the DSD of unpolarized
spins, $\mathcal{P}_{0} (L) = \langle N_{d_L}^{0} / N_{d 0} \rangle$
(where $N_{d_L}^{0}$ is the number of $0$ domains with size equal to
$L$), see again Appendix \ref{app:DSD}. The result that we have
obtained is given, in the thermodynamic limit, by
\begin{eqnarray}
  \mathcal{P}_{0} (L) &=& \frac{\lambda_{+} - 1}{\lambda_{+}^{L}} .
  \label{eq:DSDunpTh}
\end{eqnarray}
In Eqs. (\ref{eq:DSDpolTh}) and (\ref{eq:DSDunpTh}), $\lambda_+ = (1 +
e^{-\beta \delta} + \theta) / 2$.  Equations \eqref{eq:DSDpolTh} and
\eqref{eq:DSDunpTh} and their derivation are the main result of this
work. One should note that $L$ is geometrically distributed in both cases.
In Fig. \ref{fig:DSDgen}, we plot the DSD as a function of $L$ (with
logarithmic scale in the $y$-axis), for different values of $T/\gamma$
and of the ratio $\delta/\gamma$. We plot the DSD of
unpolarized spins when the edge is non-passivated [Fig. \ref{fig:DSDgen}(a)] and the
DSD of polarized spins when the edge is hydrogen-passivated
[Fig. \ref{fig:DSDgen}(b)].

The characteristic functions of these two distributions can be readily
computed from Eqs. \eqref{eq:DSDpolTh} and \eqref{eq:DSDunpTh}. We obtain
in the case of domains with polarized spins,
\begin{eqnarray}
  \hat{\mathcal{P}}_{\pm}(w) &=& \frac{\lambda_{+} e^{\beta \delta} - 1}{\lambda_{+} e^{\beta \delta} e^{-i w} - 1} 
  \,, \label{eq:CharactFuncPol}
\end{eqnarray}
while in the case of unpolarized domains, we have
\begin{eqnarray}
  \hat{\mathcal{P}}_{0}(w) &=& \frac{\lambda_{+} - 1}{\lambda_{+} e^{-i w} - 1} \,. \label{eq:CharactFuncUnp}
\end{eqnarray}

Since these distributions are geometric distributions and hence all their moments exist, their characteristic 
functions, as given by Eqs. (\ref{eq:CharactFuncPol}) and (\ref{eq:CharactFuncUnp}), are analytic at $w=0$, as can be seen by direct inspection. This result is equivalent to the statement that the distributions do not have fat tails.

The first moment of these distributions, gives us the average size of,
respectively, the domains of polarized and unpolarized spins. The
explicit expression for the average size of the domains of polarized
spins reads [see Appendix \ref{app:DSD}, Eq. (\ref{eq:MomDist})],
\begin{eqnarray}
  \bar{L}_{\pm} &=& \frac{\lambda_{+} e^{\beta \delta}}{\lambda_{+} e^{\beta \delta} - 1} \,, \label{eq:LmedExactPol}
\end{eqnarray} 
while the average size of the domains of unpolarized spins reads
\begin{eqnarray}
  \bar{L}_{0} &=& \frac{\lambda_{+}}{\lambda_{+} - 1} \,. 
  \label{eq:LmedExactUnp}
\end{eqnarray}

It is interesting to compare the results given in Eqs.
\eqref{eq:LmedExactPol} and \eqref{eq:LmedExactUnp} with the results
obtained from a different (and rather natural) definition of the
average domain size, namely $\tilde{L}_{\pm} \equiv \langle n_{pol}
\rangle / \langle n_{d \pm} \rangle$ and $\tilde{L}_{0} \equiv \langle
n_{unp} \rangle / \langle n_{d 0} \rangle$.  We obtain for
$\tilde{L}_{\pm}$, the result
\begin{eqnarray}
  \tilde{L}_{\pm} &=& \frac{ 4 e^{-2 \beta \gamma} + e^{- \beta \delta} (-1 + e^{-\beta \delta} + \theta) }
  {4 e^{-2\beta \gamma}} , 
  \label{eq:Lmed}\,.
\end{eqnarray}
Moreover, in the thermodynamic limit, $\tilde{L}_{0}$ can be written
in terms of $\tilde{L}_{\pm}$ as $\tilde{L}_{0} = \langle n_{unp}
\rangle/ \langle n_{d 0} \rangle = \big( 1 - \langle n_{pol} \rangle
\big) / \langle n_{d 0} \rangle \approx 1 / \langle n_{d 0} \rangle -
\tilde{L}_{\pm}$. If we now substitute $\lambda_+$ by its definition
in Eqs. \eqref{eq:LmedExactPol} and \eqref{eq:LmedExactUnp}, we
can show that $\bar{L}_{\pm} = \tilde{L}_{\pm}$ and $\bar{L}_{0} =
\tilde{L}_{0}$, i.e. the two definitions yield identical results. This
equality suggests that the statistical variables $N_{d_{L}}/N_{d}$ and
$N_d$ are independent in the thermodynamic limit, i.e. that the
fraction of domains with size $L$ is independent of the number of the
said domains in the limit of an infinite system. However, we were as
yet unable to prove that such independence holds in the whole
temperature range.

\begin{figure}[htp!]
  \centering
  \includegraphics[width=0.98\columnwidth]{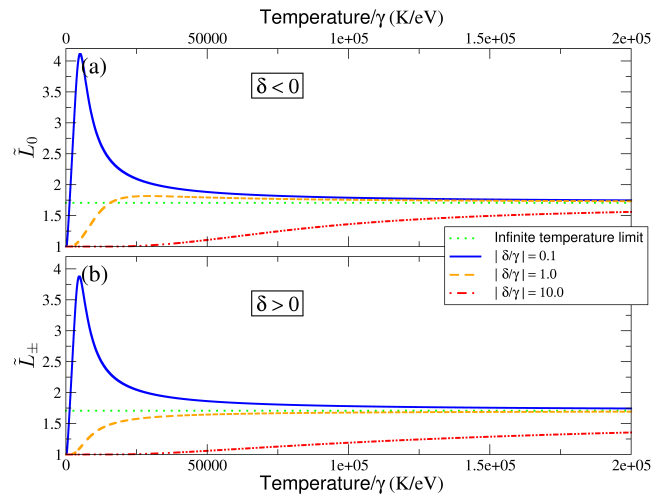}
  \caption{(Color online) Plot of the average domain size of the
    minority domains as a function of $T/\gamma$. (a) Average domain
    size of unpolarized domain sites (unpassivated edge),
    $\tilde{L}_{0}$. (b) Average domain size of polarized domain sites
    (hydrogen-passivated edge), $\tilde{L}_{\pm}$. The different
    curves in each plot, stand for three different values of the ratio
    $\delta/\gamma$. The full dark blue lines stand for $|
    \delta/\gamma | = 0.1$; the dashed orange lines stand for $|
    \delta/\gamma | = 1.0$; the dashed-dotted red lines stand for $|
    \delta/\gamma | = 10.0$. Note that the $\delta/\gamma$ ratio is
    negative for the non-passivated case (since $\delta < 0$) and
    positive for the passivated case (since $\delta > 0$). The green
    dotted flat line represents the infinite temperature limit for
    both cases, which is $1 + 1 / \sqrt{2}$.}
  \label{fig:Lmed0}
\end{figure}

In Fig. \ref{fig:Lmed0}, we plot the average domain size of the minor
domains in each of the two cases ($\bar{L}_{0}$ for non-passivated
edges and $\bar{L}_{\pm}$ for hydrogen-passivated edges - see the
previous paragraph) as a function of $T/\delta$, for selected values of
the ratio $\delta/\gamma$.

From Figs. \ref{fig:npol}-\ref{fig:Lmed0}, we confirm that
irrespective of the value of the dimensionless temperature $T/\gamma$,
the system always presents a finite concentration of defects at any
finite temperature, as is to be expected for a one-dimensional system
with short-range interactions. However, since our model does not
possess the full $\mathbf{Z}_3$ symmetry characteristic of a true Potts-like
model (in which case the Peierls argument does apply), the results
obtained are qualitatively different from those obtained for an Ising
chain, where the formation of domains of macroscopic size fully
destroys order at any $T \neq 0$ (in our case, the energy of formation
of a domain does depend on the domain's size). In contrast with what
happens in the one-dimensional Ising model, here the disorder tends to
zero with decreasing temperature, being exactly zero only at
$T=0$. Depending on the exchange parameters of the system, the
concentration of minority domains at room temperature (and hence the
degree of disorder of the edge at such temperature) may or may not
present a large value. In addition, the smaller the ratio
$\delta/\gamma$ is, the less stable the edge is to the effect of
thermal disorder. As expected, the average mean size of
the minority domains increases with temperature. Finally, the larger
the ratio $\delta/\gamma$ is, the larger is the size of the minority
domains. This is to be expected, so as to minimize the number of
domain walls (links 0+ and 0-, whose exchange parameter is $\gamma$)
relatively to the number of +- links (whose exchange parameter is
$\delta$).

In Appendix \ref{sec:ExchangeParamsResults}, based on {\it ab-initio}
calculations that we have both performed ourselves (case
$\mathcal{C}_{1}$) and that we have obtained from the existing
literature (case $\mathcal{C}_{2}$), we compute specific values of the
exchange parameters of the model that we have introduced, $(
\delta_{\mathcal{C}_{1}}, \gamma_{\mathcal{C}_{1}} )$ and $(
\delta_{\mathcal{C}_{2}}, \gamma_{\mathcal{C}_{2}} )$.  Using these
particular values of the exchange parameters, we present there plots 
of the thermodynamic quantities introduced in Eqs. (\ref{eq:polDensity})-(\ref{eq:Lmed}),
as functions of the absolute temperature, rather than presenting them
as functions of the reduced temperature $T/\gamma$ and the ratio
$\delta/\gamma$. From these results, one can conclude that, in both
cases $\mathcal{C}_{1}$ and $\mathcal{C}_{2}$, the exchange
parameters calculated from the {\it ab-initio} are such that the
ground-state edge configuration is robust with respect to the effect
of thermal disorder.

Nevertheless, it should also be stated that due to the sensitivity of
the system's thermodynamic behavior on the precise numerical value of
the exchange parameters (see Appendix \ref{sec:ExchangeParamsResults}), 
this conclusion may well be challenged in the future, in case more detailed 
{\it ab-initio} calculations yield different numerical values for the exchange
parameters. We should emphasize in this regard that the {\it ab-initio} calculations 
that we have performed were done using narrow ribbons, where an 
interaction between the two edges of the ribbon can be observed. Moreover, 
these calculations did not take into account either the
spin-polarization of electrons on the edge, or the relaxation of the
atoms along the transverse direction of the ribbon and that such
complications need to be addressed in future publications.

Note that our model is necessarily an oversimplified one. Firstly, it
assumes that the state of passivation of the edge is a quenched variable 
determined by the concentration of $H_2$ molecules present in the atmosphere of
the experiment. This is not an entirely realistic assumption, since it
is to be expected that the binding and unbinding of $H$ atoms to an
hexagon or heptagon (pentagons have no dangling-bonds to which $H$
atoms can bind to) is influenced by temperature and
pressure. Taking such observation into account in our model would
imply the introduction of a chemical potential regulating the chemical
equilibrium between the passivating atoms attached to the edge and
those in the atmosphere surrounding the ribbon.  Moreover, in the most
general case, one would also have to allow the state of passivation of
an hexagon or heptagon to be a statistical variable, since the bare
exchange parameters between neighboring sites should depend on their
state of passivation. This would imply the introduction of a
Potts-like model with a higher number of colors (corresponding to both
hydrogen-passivated and non-passivated edge polygons).

Finally, it is to be expected that in an experiment, the edge may be
passivated by other atomic or molecular species present in the gaseous
environment surrounding the ribbon (namely oxygen, nitrogen, water,
etc.) and not just by hydrogen. In order to take into account the
presence of competing species, one would need to consider a Potts-like
model with a yet higher number of colors, together with additional
exchange parameters associated with the interaction between different
kinds of passivation between neighboring polygons, each of which would
need to be computed from {\it ab-initio} simulations. Moreover, one
would have to introduce a chemical potential for each species,
regulating the chemical equilibrium between the passivating atoms of
that species attached to the edge and those in the atmosphere
surrounding the ribbon. This would make the model increasingly
difficult to study using a simple analytical approach as the one
presented above.

\section{Conclusion}

\label{secCon}

In this work, we have treated the zigzag edge (in the presence of SW carbon rings) of 
a graphene ribbon as a one-dimensional system and introduced a three-color Potts-like model 
to study its thermodynamic properties regarding the presence of thermal disorder. We have
  shown how to extract the effective parameters that describe the
  model from {\it ab-initio} calculations and how to use these
  numerical values to determine the temperature dependence of the
  defect size and defect concentration.

  As is to be expected for a one-dimensional system with short-range
  interactions, we concluded that
  the edge is always disordered at any finite temperature. More
  importantly, this model allowed us to make quantitative predictions
  for the concentration and size of the defective domains at a given
  temperature, both for the totally passivated and for the totally
  non-passivated zigzag edge. The defect concentration was found to 
  be exponentially dependent on the exchange parameters of the model. 
  Depending on the actual value of these parameters, the concentration of 
  defects can become quite large at room temperatures, and may thus have 
  a significant effect on the conductivity of the zigzag ribbon.
  We have also computed the DSD for the totally passivated and non-passivated
  edge and have concluded that these distributions do not have fat-tails.

  Edge disorder may strongly influence the conductance of graphene-based
  devices. However, the equilibrium mechanisms described in this
  work only place a lower bound on the quantity of disorder present
  at the edges of graphene nanostructures. Equivalently, they put an
  upper limit in the value of the conductance that can be obtained in
  these devices.\cite{Lewenkopf_PRB:2008, Mucciolo_PRB:2009}

  \begin{acknowledgements}

    We acknowledge helpful discussions with N. Peres, J. Lopes dos
    Santos, P. Ribeiro, E. Lage, R. Ribeiro and
    A. L\"{a}uchli. J. N. B. R. was supported by the Portuguese
    Foundation for Science and Technology (FCT) through Grant
    No. SFRH/BD/44456/2008. J.E.S. acknowledges support by the
    Visitors Program of the MPIPkS and by the MPICPfS at the early
    stages of this work. J.E.S. work contract is financed in the
    framework of the Program of Recruitment of Post Doctoral
    Researchers for the Portuguese Scientific and Technological
    System, within the Operational Program Human Potential (POPH) of
    the QREN, participated by the European Social Fund (ESF) and
    national funds of the Portuguese Ministry of Education and Science
    (MEC). He also acknowledges support provided to the current
    research project by FEDER through the COMPETE Program and by FCT
    in the framework of the Strategic Project PEST-C/FIS/UI607/2011.
    A. H. C. N. acknowledges DOE grant DE-FG02-08ER46512, ONR grant
    MURI N00014-09-1-1063, and the NRF-CRP award "Novel 2D materials
    with tailored properties: beyond graphene" (R-144-000-295-281).

  \end{acknowledgements}

\appendix

\section{The exchange parameters from {\it ab-initio} results}

\label{sec:ExchangeParams}

In this appendix we will show how one can compute the exchange parameters 
of the Potts-like model, from {\it ab-initio} results of zigzag ribbons
with reconstructed edges.

The energy per edge atom of periodic edge configurations such as $zz$, $zz(57)$, $zz(576)$ 
and $zz(576^{n})$ (where $n$ stands for the number of hexagons in the periodic edge configuration 
- see Fig. \ref{fig:reczag_family}), can be computed either using the three-color 
Potts-like model proposed in the main text (for particular values of the parameters $\gamma$ and 
$\delta$), or using the {\it ab-initio} results for the edge energies computed from
density functional theory. From a least squares method, we can then
compute the exchange parameters, $\gamma$ and $\delta$, of the
Potts-like model, in such a way that the latter describes, to a good
degree of accuracy, the {\it ab-initio} results.

The edge energies (per unit cell of the perfect zigzag edge) of different periodic 
reconstructions of the edge, for example, $zz(57)$, $zz(576)$, $zz(5766)$, etc., 
are given, in the scope of the previously introduced Potts-like model, by
\begin{subequations}
  \begin{eqnarray}
    E\big(zz(57)\big) &=& J_{+-} , \\
    E\big(zz(576^{n})\big) &=& \frac{J_{+-} + 2 J_{+0} + (n-1) J_{00}}{n+2} , \\
    E\big(zz\big) &=& J_{00} ,
  \end{eqnarray}
\end{subequations}
where $n$ stands for the number of edge hexagons present in a unit
cell. Expressing these energies relative to the clean edge energy
$\Delta E\big( zz(576^{n}) \big) = E\big(zz(576^{n})\big) -
E\big(zz\big)$, with the latter set to zero (i.e. $J_{00}=0$, as
above), we obtain,
\begin{subequations} \label{eqDelta}
  \begin{eqnarray}
    \Delta E\big( zz(57); \delta, \gamma \big) &=& \delta, \\
    \Delta E\big( zz(576^{n}); \delta, \gamma \big) &=& \frac{\delta + 2 \gamma}{n+2},
  \end{eqnarray}
\end{subequations}
where we made the substitutions $J_{+-} = \delta$ and $J_{+0} =
\gamma$.

\begin{figure}[htp!]
  \centering
  \includegraphics[width=0.98\columnwidth]{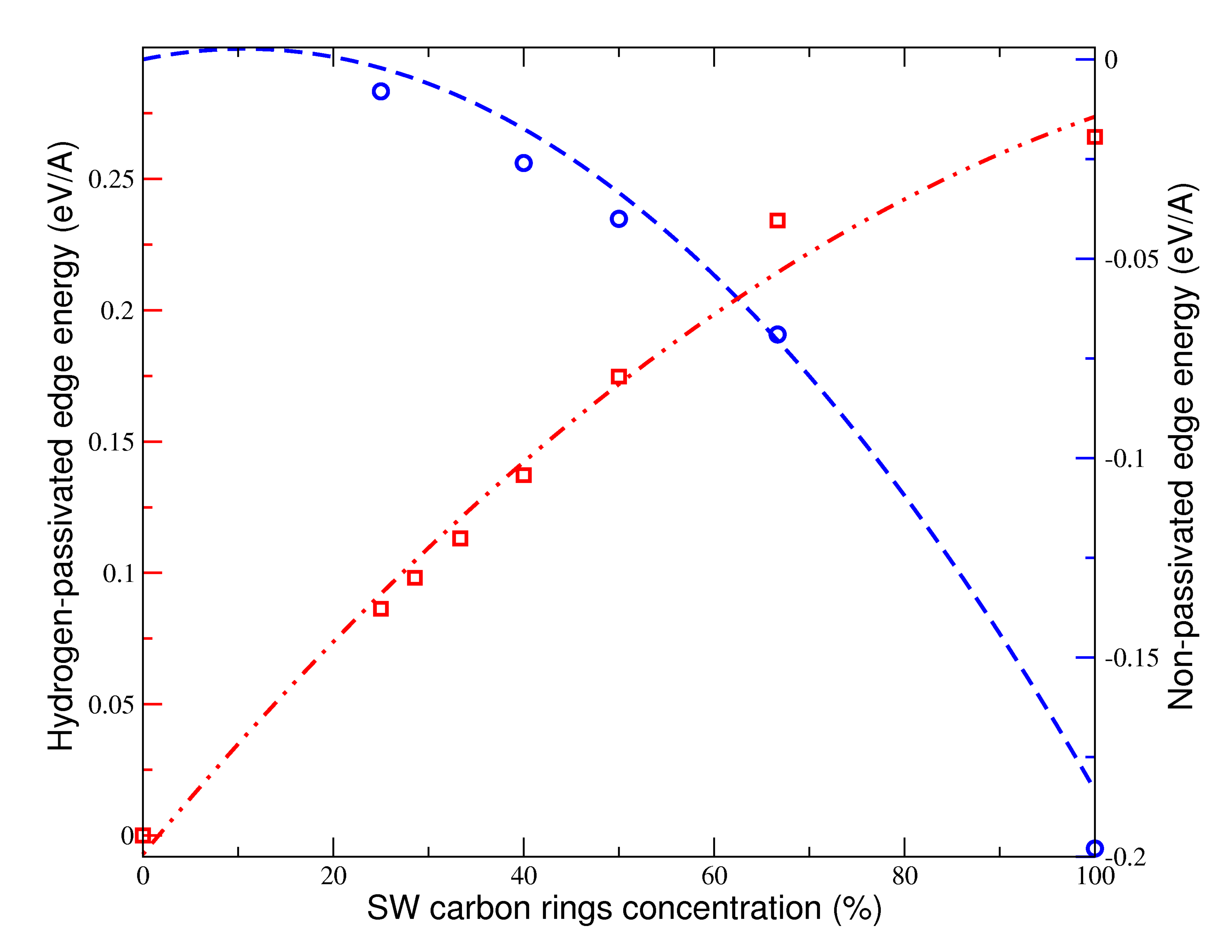}
  \caption{(Color online) Energies of the partially reconstructed
    edges, measured relative to the pristine $zz$ edge, as a function
    of the SW carbon rings concentration. Dots represent the edge
    energies obtained from {\it ab-initio} calculations, while dashed
    lines correspond to a polynomial interpolation of the results
    obtained from the Potts-like model using the exchange parameters
    computed with the least squares method. The red squares and red
    dashed-dotted line represent the results obtained for
    hydrogen-passivated zigzag ribbons (left $y$-axis), whereas the
    blue circles and blue dashed line represent the edge energies of
    non-passivated zigzag ribbons (right $y$-axis).}
  \label{fig:LDAresults}
\end{figure}

We now consider the energy $\epsilon_{n}$ of the edge $zz(576^{n})$
referred to the pristine zigzag edge, as obtained from {\it ab-initio}
calculations (see Fig. \ref{fig:LDAresults}). The exchange parameters
$\gamma$ and $\delta$ can be obtained from a minimization of the sum
of the squared differences between $\Delta E\big(zz(576^{n})\big)$, as
given by Eqs. (\ref{eqDelta}), and $\epsilon_{n}$,
\begin{eqnarray}
  S(\delta, \gamma) &=& \sum_{n=0} \Big[ \Delta E\big( zz(576^{n}); \delta, \gamma \big) - \epsilon_{n} \Big]^{2} .
\end{eqnarray}
The uncertainty on the computed exchange parameters, is given by
\begin{eqnarray}
  \sigma_{z} = \sum_{n=0} \Big[ \sigma_{n}^{2} \Big( \frac{\partial z}{\partial \epsilon_{n}} \Big)^{2} \Big] ,
\end{eqnarray}
where $z$ stands for $\gamma$ or $\delta$, whose expression as a
function of $\epsilon_{n}$ is computed from minimization of
$S(\delta,\gamma)$, and where $\sigma_{n}$ are the uncertainties in
the {\it ab-initio} energies $\epsilon_{n}$.



\label{sec:ExchangeParamsResults}

In order to obtain indicative values for the exchange parameters of
the three-color Potts-like model introduced in the text, we have both used
{\it ab-initio} results on non-passivated zigzag edges already
published in the literature,\cite{Huang_PRL:2009} and have ourselves
performed {\it ab-initio} calculations on hydrogen-passivated
edges. In Fig. \ref{fig:LDAresults}, we plot the edge energies
(relative to the energy of the pristine zigzag edge), obtained from
{\it ab-initio} calculations of edge reconstructed zigzag ribbons with
both hydrogen-passivated edges and non-passivated edges.

From the {\it ab-initio} results summarized in Fig. \ref{fig:LDAresults}, and after employing 
the method just described to compute the exchange parameters in both cases, we obtain the 
following values for the exchange parameters:

\begin{itemize}
\item Hydrogen-passivated edge (case $\mathcal{C}_{1}$): We have
  performed {\it ab-initio} calculations for hydrogen-passivated
  ribbons with various concentrations of SW carbon rings (see Fig.
  \ref{fig:LDAresults}).\footnote{The Density Functional Theory (DFT)
    calculations were performed using the code
    AIMPRO,\cite{Rayson_CPC:2008} under the Local Density
    Approximation (LDA).  The Brillouin-zone (BZ) was sampled for
    integrations according to the scheme proposed by
    Monkhorst-Pack.\cite{Monkhorst_PRB:1976} The core states were
    accounted for by using the dual-space separable pseudo-potentials
    by Hartwigsen, Goedecker, and Hutter.\cite{Hartwigsen_PRB:1998}
    The valence states were expanded over a set of $s$-, $p$-, and
    $d$-like Cartesian-Gaussian Bloch atom-centered functions. The
    total energies in the self-consistency cycle were converged such
    that changes in energy between two iterations and the
    electrostatic energy associated with the difference between input
    and output charge densities were both less than $2.7 \times
    10^{-4}$ eV. The k-point sampling ranged from $12 \times 4 \times
    1$ for the $zz(57)$ edge to $4 \times 4 \times 1$ for the
    $zz(57666666)$ edge and the atoms were relaxed in order to find
    their equilibrium positions. The ribbons were simulated within a
    supercell geometry using vacuum layers of $12.7 \mathring{A}$ in
    the ribbon plane and $10.6 \mathring{A}$ in the normal direction
    in order to avoid interactions between ribbons in adjacent cells.}
  We have assumed the same value $\sigma_{n} = 0.01 \, e\mbox{V}$ for
  all uncertainties.\footnote{We have estimated the uncertainty
    associated with our {\it ab-initio} calculations of the edge
    energy per angstrom (for every SW carbon ring periodicity at the edges),
    to be given by $\sigma' \approx 0.01 e\mbox{V}$ per unit cell of
    pristine edge.}  The values obtained for the parameters were
  $\gamma_{\mathcal{C}_{1}} = (0.53 \pm 0.03) \, e\mbox{V}$ and
  $\delta_{\mathcal{C}_{1}} = (0.65 \pm 0.04) \, e\mbox{V}$.
\item Non-passivated edge (case $\mathcal{C}_{2}$): The {\it
    ab-initio} results used in this case were extracted from the work
  of Huang {\it et al.}.\cite{Huang_PRL:2009} We have assumed the same
  value $\sigma_{n} = 0.01 \, eV$ for all uncertainties.\footnote{From
    Figure $5(a)$ of Huang et al.,\cite{Huang_PRL:2009} we have
    estimated the uncertainty associated with the edge energies per
    unit of length to be $\sigma' = 0.004
    e\mbox{V}/\mathring{\mbox{A}}$. As in this work we are using units
    of energy per unit cell of the pristine edge, $\sigma = \sigma'
    \times 1.42 \sqrt{3} \approx 0.01 e\mbox{V}$.} The values obtained
  for the parameters were $\gamma_{\mathcal{C}_{2}} = (0.03 \pm 0.02)
  \, e\mbox{V}$ and $\delta_{\mathcal{C}_{2}} = (- 0.49 \pm 0.03) \,
  e\mbox{V}$.
\end{itemize}

In Fig. \ref{fig:LDAresults}, we plot the results obtained from {\it
  ab-initio} calculations (isolated dots), compared with the
polynomial interpolation of these results (dashed-lines), which was
obtained from the least squares method, as described above.  The fact that these 
curves are in good agreement with the {\it ab-initio} results justifies {\it a
  posteriori} the use of a Potts-like model with only nearest-neighbor
interactions.

\begin{figure}[htp!]
  \centering
  \includegraphics[width=0.98\columnwidth]{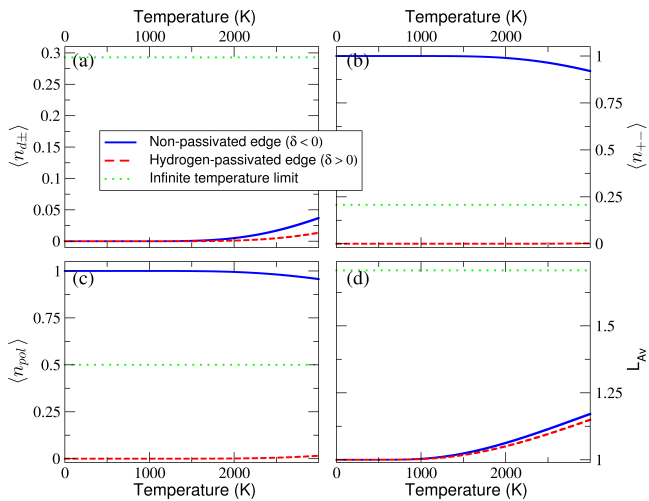}
  \caption{(Color online) Plot of the four thermodynamic quantities introduced in the
    main text, for the values of the exchange parameters obtained from
    the ab-initio results: $\delta_{\mathcal{C}_{1}} = 0.66$ and
    $\gamma_{\mathcal{C}_{1}} = 0.52$ for the hydrogen-passivated edge
    (red dashed curves); $\delta_{\mathcal{C}_{2}} = -0.49$ and
    $\gamma_{\mathcal{C}_{2}} = 0.03$ for the non-passivated edge
    (blue full curves). Panel (a) shows the concentration of polarized
    domains, $\langle n_{d \pm} \rangle$. Panel (b) shows the concentration 
    of links between polarized spins, $\langle n_{+-} \rangle$. Panel (c) 
    shows the concentration of polarized spins, $\langle n_{pol} \rangle$. Panel
    (d) shows the average minor domain size, $L_{Av}$. The dotted green flat
    lines represent the infinite temperature limit of each quantity.}
  \label{fig:AllFour}
\end{figure}

Using the above values for the exchange parameters, the concentration of
polarized spins, $\langle n_{pol} \rangle$, defined in Eq. (\ref{eq3a}), the concentration
of links between polarized spins, $\langle n_{+-} \rangle$, defined in
Eq. (\ref{eq3b}), the concentration of polarized domains, $\langle n_{d \pm} \rangle$, defined
in Eq. (\ref{eq4}) and the average domain size of the minor domains,
$L_{Av}$, defined in Eq. (\ref{eq:Lmed}), acquire the form presented
in Fig. \ref{fig:AllFour}.

Using the exchange parameters calculated above, one concludes that the
concentration of polarized domains in the non-passivated case at room
temperature has a value $\langle n_{d} \rangle \simeq 6.65 \times
10^{-18}$ defects per unit cell of pristine zigzag edge (or $\langle
n_{d} \rangle \simeq 5.41 \times 10^{-8}$ defects per
meter).\footnote{Note that, in the thermodynamic limit, the number of
  polarized domains is equal to the number of unpolarized domains, and
  consequently, the corresponding domain densities are also equal.} In
addition, at room temperature, the unpolarized domains have an average
domain size of $\bar{L}_{0}^{MF} \simeq 1$ unit cells (or
$\bar{L}_{0}^{MF} \simeq 1.23 \times 10^{-10} m$). These small
unpolarized domains are on average $1.85 \times 10^{7} m$ apart from
each other.

In the case of a hydrogen-passivated edge, the above calculated
parameters give, at room temperature, a concentration of polarized domains
of $\langle n_{d} \rangle \simeq 1.50 \times 10^{-22}$ per unit cell
of pristine zigzag edge (or $\langle n_{d} \rangle \simeq 1.22 \times
10^{-14}$ defects per meter). The polarized domains have a mean size
of $\bar{L}_{\pm}^{MF} \simeq 1$ unit cells (or $\bar{L}_{\pm}^{MF}
\simeq 1.23 \times 10^{-10} m$). The small domains of polarized spins
are on average $8.20 \times 10^{13} m$ apart from each other.

These results show that, with the given set of effective parameters
as computed from the {\it ab-initio} methods, the ground-states in 
both the non-passivated case (totally reconstructed edge) and the 
hydrogen-passivated case (pristine zigzag edge) are very stable with 
respect to the effect of thermal disorder.

However, we should note that in our model, the sensitivity of the
thermodynamic quantities on the values of the exchange parameters is
large. In order to illustrate such fact, consider for instance that
the exchange parameters were reduced to $1/4$ of the values which we
determined above. This would imply that the room temperature
concentration of defects would be increased by several orders of
magnitude: in the non-passivated case, it would be increased to
$\langle n_{d} \rangle \simeq 7.01 \times 10^{5}$ defects per meter
(the average distance between neighboring defects would be $1.43 \mu
m$); in the hydrogen-passivated case, the defect concentration would
be increased to $\langle n_{d} \rangle \simeq 4.82 \times 10^{4}$
defects per meter (the average distance between neighboring defects
would be $20.8 \mu m$ apart). Consequently, if more detailed {\it
  ab-initio} calculations were to give significantly smaller exchange
parameters, this would imply that the edge ground-state would be much
less robust to the effect of thermal disorder.

\section{Correlation functions of the Potts-like model}
\label{appA}
The computation of correlation functions of a one-dimensional
Potts-model in a periodic system involves the computation of the trace
of a string of operators.\cite{Hintermann_HPA:1969}  For instance,
the magnetization of the system can be written, using the cyclic
invariance of the trace, as
\begin{eqnarray}
  \label{eqA1}
  \langle\, \sigma_i
  \,\rangle&=&\frac{1}{Z_{2N}}\,\sum_{\{\sigma\}}\,\mathbf{T}_{\sigma_1\sigma_2}\,\ldots
  \mathbf{T}_{\sigma_{i-1}\sigma_{i}}\,\sigma_i\,\mathbf{T}_{\sigma_{i}\sigma_{i+1}}\,
  \ldots\mathbf{T}_{\sigma_{2N}\sigma_1}\nonumber\\
  &=&\frac{1}{Z_{2N}}\,\mbox{Tr}(\hat{\sigma}\,\mathbf{T}^{2N})\,,
\end{eqnarray}
where $\mathbf{T}_{\alpha\beta}$ are the individual matrix elements of
the transfer matrix in Eq. (\ref{eq1}), while
$Z_{2N}=\mbox{Tr}(\mathbf{T}^{2N})=\lambda_+^{2N}+\lambda_0^{2N}+\lambda_-^{2N}$
is the partition function of the model and $\hat{\sigma}$ is the
$3\times 3$ matrix
\begin{equation}
  \label{eqA2}
  \hat{\sigma}=\left( \begin{array}{ccc}
      1 & 0 & 0 \\
      0& 0 & 0 \\
      0 & 0 & -1
    \end{array} \right)\enspace\,.
\end{equation}

Eq. (\ref{eqA1}) shows that the magnetization is
space-independent. Since the trace in Eq. (\ref{eqA1}) is independent
of the basis used for its calculation, we choose the one that
diagonalizes $\mathbf{T}$,
\begin{equation}
  \label{eqA3}
  \vert \lambda_0 \rangle =\frac{1}{\sqrt{2}}
  \left( \begin{array}{c}
      1\\
      0\\
      -1
    \end{array} \right)\enspace\,,
\end{equation}
which is the eigenvector corresponding to the eigenvalue
$\lambda_0=-e^{-\beta\delta}$ and
\begin{equation}
  \label{eqA4}
  \vert \lambda_\pm \rangle=
  \left( \begin{array}{c}
      \alpha_{\pm}\\
      \beta_{\pm}\\
      \alpha_{\pm}
    \end{array} \right)\enspace\,,
\end{equation}
which are the eigenvectors corresponding to the eigenvalues
$\lambda_\pm=\frac{1}{2}\left[1+e^{-\beta\delta}\pm\sqrt{
    (1-e^{-\beta\delta})^2+8\,e^{-2\beta\gamma}}\right]$, where
$\alpha_{\pm}$ is given by
\begin{equation}
  \label{eqA5}
  \alpha_{\pm}=\frac{\lambda_{\pm}-1}{\sqrt{2}\,[(\lambda_{\pm}-1)^2+2\,e^{-2\beta\gamma})]^{1/2}}\,,
\end{equation}
and where $\beta_{\pm}^2=1-2\alpha_{\pm}^2$ (normalization
condition). It can be easily checked that these three vectors form an
orthonormal basis. Expressing the trace in terms of this basis, one
obtains for the magnetization Eq. (\ref{eqA1}), the result
\begin{equation}
  \label{eqA6}
  \langle\, \sigma_i
  \,\rangle=\frac{1}{Z_{2N}}\sum_{\mu=0,\pm 1}\,\lambda_\mu^{2N}\,\langle \lambda_\mu \vert \,\hat{\sigma}\,\vert \lambda_{\mu} \rangle=0\,,
\end{equation}
since $\langle \lambda_\mu \vert \,\hat{\sigma}\,\vert \lambda_{\mu} \rangle=0$ for
each one of the eigenvectors of $\mathbf{T}$. This equality merely
reflects the symmetry of the model with respect to an interchange of
$+$ with $-$ spins that is present by construction. In order to infer
the existence of a phase transition at $T=0$ in the absence of a
(infinitesimal) field that explicitly breaks this symmetry, one needs
to consider the behavior of higher-order correlation functions.

The spin-spin correlation function
$\langle\,\sigma_i\sigma_{i+j}\,\rangle$ is given by
\begin{eqnarray}
  \label{eqA7}
  \langle\, \sigma_i\sigma_{i+j}\,\rangle
  &=&\frac{1}{Z_{2N}}\,\mbox{Tr}(\mathbf{T}^{2N-j}\hat{\sigma}\,\mathbf{T}^{j}\hat{\sigma})\nonumber\\
  &=&\frac{1}{Z_{2N}}\sum_{\mu,\nu}\lambda_\mu^{2N-j}
  \lambda_\nu^{j}\mid \langle \lambda_\mu \vert \,\hat{\sigma}\,\vert \lambda_{\nu} \rangle \mid^2,
\end{eqnarray}
where we have used a representation of the unit-operator in terms of
the eigenstates of $\mathbf{T}$, on going from the first to the second
line of Eq. (\ref{eqA7}). At $T\neq 0$ and in the thermodynamic limit
$N\rightarrow\infty$, the only term in the numerator of
Eq. (\ref{eqA7}) that survives, is the one with $\nu=0$, $\mu=+1$ and
$Z_{2N}\approx \lambda_+^{2N}$.  Thus, we obtain in this case, since
$\langle \lambda_+ \vert \,\hat{\sigma}\,\vert \lambda_{0} \rangle=\sqrt{2}\alpha_+$,
\begin{equation}
  \label{eqA8}
  \langle\, \sigma_i\sigma_{i+j}\,\rangle=2\alpha_+^2\,\left(\frac{\lambda_0}{\lambda_+}\right)^j\,.
\end{equation}
If $T\neq 0$, $\langle\, \sigma_i\sigma_{i+j}\,\rangle\rightarrow 0$
if $j\rightarrow\infty$, showing that the magnetization of the system
is zero at any finite temperature, as is to be expected for any system
with $\mathbb{Z}_2$ symmetry in 1d. At $T=0$, one has to distinguish
three cases: $\delta>0$, in which case $\lambda_+\rightarrow 1$ and
both $\lambda_0$ and $\lambda_-$ go to zero. In that case,
Eq. (\ref{eqA7}) still holds and the ground-state is simply the
$0000\ldots$ state, with no associated magnetization.  If, on the
other hand $\delta<0$, $\lambda_+\rightarrow \infty$,
$\lambda_0\rightarrow -\infty$ and $\lambda_-\rightarrow 0$. In that
case, one has to consider again Eq. (\ref{eqA6}), since the terms
$\langle \lambda_+ \vert \,\hat{\sigma}\,\vert \lambda_{0} \rangle$ and
$\langle \lambda_0 \vert \,\hat{\sigma}\,\vert \lambda_+ \rangle$ contribute equally to
it. Thus, we obtain $\langle\, \sigma_i\sigma_{i+j}\,\rangle=(-1)^j$,
which shows that the anti-ferromagnetic states '$\ldots +-+- \ldots$'
and '$\ldots -+-+ \ldots$ are the two degenerate ground-states. In
this case, the system shows a transition to a finite (staggered)
magnetization at zero temperature. Finally, if $\delta=0$,
$\lambda_\pm\rightarrow 1$, $\lambda_0\rightarrow -1$ and all terms
$\langle \lambda_\pm \vert \,\hat{\sigma}\,\vert \lambda_{0} \rangle$ and
$\langle \lambda_0 \vert \,\hat{\sigma}\,\vert \lambda_\pm \rangle$ contribute to
Eq. (\ref{eqA7}). We obtain $\langle\,
\sigma_i\sigma_{i+j}\,\rangle=\frac{2}{3}\,(-1)^j$, which shows that
there are three degenerate ground-states ``$\ldots 0000 \ldots$'',
``$\ldots +-+- \ldots$'' and ``$\ldots -+-+ \ldots$''. One can also show
that a phase transition is present when $\delta\leq 0$, if one writes
Eq. (\ref{eqA8}) as $\langle\,
\sigma_i\sigma_{i+j}\,\rangle=2\alpha_+^2\,(-1)^j\,e^{-j/\xi}$, where
$\xi=1/\ln(\lambda_+/\mid \lambda_0\mid)$ is the correlation length of
the model.  If $\delta>0$, $\xi=0$ at $T=0$ and no phase transition
occurs, but if $\delta\leq 0$, $\xi\rightarrow\infty$ at $T=0$
indicating the presence of a phase transition. Note that the presence
of a phase transition has at most a marginal effect on the results
presented in the main text, since such a phase transition is due to
the existence of a $\mathbb{Z}_2$ symmetry in the model, whereas the
formation of minority domains of either $0$'s or $+-$ relies on states
not related by such a symmetry.

One can also use the transfer matrix formalism to compute the
probability $\langle\, \delta_{\sigma_i\sigma_{i+j},-1}\,\rangle$ that
the spins at sites $i$ and $i+j$ are anti-parallel. One uses the
identity $\delta_{\sigma_i\sigma_{i+j},-1}=
\frac{1}{2}\,\sigma_i\sigma_{i+j}(\sigma_i\sigma_{i+j}-1)$, which can
be easily checked by substituting $\sigma_i$ and $\sigma_{i+j}$ by
their values $0,\pm 1$. Since we have already computed the spin-spin
correlation function above, we are left with the computation of
$\langle\,\sigma_i^2\sigma_{i+j}^2\,\rangle$.  Following the same
steps as above, we obtain
\begin{equation}
  \label{eqA9}
  \langle\, \sigma_i^2\sigma_{i+j}^2\,\rangle
  =\frac{1}{Z_{2N}}\sum_{\mu,\nu}\lambda_\mu^{2N-j}
  \lambda_\nu^{j}\mid\langle\lambda_\mu \vert \,\hat{\sigma}^2\,\vert \lambda_{\nu} \rangle\mid^2,
\end{equation}
where $\hat{\sigma}^2$ is the matrix
\begin{equation}
  \label{eqA10}
  \hat{\sigma}^2=\left( \begin{array}{ccc}
      1 & 0 & 0 \\
      0& 0 & 0 \\
      0 & 0 & 1
    \end{array} \right)\enspace\,.
\end{equation}
At $T\neq 0$, the only terms that need to be considered in
Eq. (\ref{eqA9}) are those involving
$\langle \lambda_+ \vert \,\hat{\sigma}^2\,\vert \lambda_+ \rangle$ and
$\langle \lambda_+ \vert \,\hat{\sigma}^2\,\vert \lambda_0 \rangle$.  Taking this into
account, as well as the expression Eq. (\ref{eqA8}) for $\langle\,
\sigma_i\sigma_{i+j}\,\rangle$, one finally obtains for
$\langle\,\delta_{\sigma_i\sigma_{i+j},-1}\,\rangle$, the result
\begin{eqnarray}
  \label{eqA11}
  \langle\,\delta_{\sigma_i\sigma_{i+j},-1}\,\rangle&=&2\alpha_+^4 +2\alpha_+^2\alpha_-^2\,\left(
    \frac{\lambda_0}{\lambda_+}\right)^j
  \nonumber\\
  &&\mbox{}-\alpha_+^2\,\left(\frac{\lambda_0}{\lambda_+}\right)^j\,.
\end{eqnarray}

Using Eq. (\ref{eqA8}) with $j=0$ and Eq. (\ref{eqA11}) with $j=1$ in
the expression for $n_d=\langle\,\sigma_i^2\,\rangle-
\langle\,\delta_{\sigma_i\sigma_{i+1},-1}\,\rangle$ given in Section
\ref{sec:EdgeThermo}, we obtain
$n_d=2\alpha_+^2(1-e^{-\beta\delta}/\lambda_+)$, which is exactly the
result given in Eq. (\ref{eq5}).

\section{The Domain Size Distribution}
\label{app:DSD}

In what follows, we will compute the DSD of
the $0$-spins domains and of $\pm$-spin domains. Throughout the
computation, we will assume PBCs for the system. We will illustrate
the computation of the DSD of $0$-spins, pinpointing
the differences with the computation of the DSD of
$\pm$-spins.

In the context of the exact calculation of the DSD, the thermal
average of the distribution of the sizes of domains of $0$-spins is
defined by
\begin{eqnarray}
  \bar{L}_0 &=& \sum_{L=1}^{2 N} L\,P_0(L) , \label{eq:definingPLunp}
\end{eqnarray}
where $P_0(L)$ is the domain size distribution of domains of $0$-spins
(thermal average of the fraction of domains of size $L$). This
quantity is given by
\begin{eqnarray}
  P_{0}(L) &=& \Bigg\langle \frac{N_{d_{L}}^0}{N_{d 0}} \Bigg\rangle , \label{eq:DSDdefunp}
\end{eqnarray}
where $N_{d_L}^0$ stands for the number of domains of $0$-spins with
size $L$, while $N_{d 0}$ stands for the total number of $0$-spins
domains regardless of their size. An analogous expression can be
written for the DSD of $\pm$-spins.

We start by defining the operator that verifies in every possible way
if the spin $i$ is in a domain of spins $0$ with size $L$,
\begin{eqnarray}
  f_{i, L}^{0} &\equiv& \sum_{k=0}^{L-1} \Bigg[ \sigma_{i-k-1}^{2} \bigg( \prod_{\gamma=0}^{L-1} 
  \big( 1 - \sigma_{i-k+\gamma}^{2} \big) \bigg) \sigma_{i-k+L}^{2} \Bigg] . \nonumber \\ \label{eq:f0i1}
\end{eqnarray}

The above definition is valid for cases where the domain has a size $L
\leq 2 N - 2$, where $2 N$ stands for the total number of spins in the
one-dimensional chain. In the case where $L = 2 N-1$ and $L=2 N$, this
definition is modified. It reads
\begin{eqnarray}
  f_{i, L=2 N-1}^{0} &\equiv& \sum_{k=0}^{2 N-2} \Bigg[ \sigma_{i-k-1}^{2} \prod_{\gamma=0}^{2 N-2} 
  \big( 1 - \sigma_{i-k+\gamma}^{2} \Big)\Bigg] , \label{eq:f0i2} \\
  f_{i, L=2 N}^{0} &\equiv& \prod_{\gamma=0}^{2 N-1} \big( 1 - \sigma_{i+\gamma}^{2} \big) . \label{eq:f0i3}
\end{eqnarray}

We can analogously define an operator verifying in every possible way
if the spin $i$ is in a domain of spins $\pm$ with size $L$, namely
$f_{i, L}^{\pm}$. To do this, it suffices to substitute, in Eqs.
(\ref{eq:f0i1})-(\ref{eq:f0i3}), the operators $\sigma^2$ by $1 -
\sigma^2$. For $L \leq 2 N - 2$, $f_{i, L}^{\pm}$ reads
\begin{eqnarray}
  f_{i, L}^{\pm} &\equiv& \sum_{k=0}^{L-1} \Bigg[ \big( 1 - \sigma_{i-k-1}^{2} \big) \prod_{\gamma=0}^{L-1} 
  \sigma_{i-k+\gamma}^{2} \big( 1 - \sigma_{i-k+L}^{2} \big) \Bigg] , \nonumber \\ \label{eq:fpmi1}
\end{eqnarray}
whereas, for $L = 2 N -1$ and $L = 2 N$, we have
\begin{eqnarray}
  f_{i, L=2 N-1}^{\pm} &\equiv& \sum_{k=0}^{2 N-2} \Bigg[ \big( 1 - \sigma_{i-k-1}^{2} \big) 
  \prod_{\gamma=0}^{2 N-2} \sigma_{i-k+\gamma}^{2} \Bigg] , \label{eq:fpmi2} \\
  f_{i, L=2 N}^{\pm} &\equiv& \prod_{\gamma=0}^{2 N-1} \sigma_{i+\gamma}^{2} . \label{eq:fpmi3}
\end{eqnarray}

Given this, if for a given configuration of the edge, we want to count
the number of unpolarized spins in domains of size $L$, $N_{L}^0$, we
just need to perform the sum of the operators $f_{i , L}^{0}$ over
every site in the one-dimensional chain. Explicitly, it reads
\begin{eqnarray}
  N^0_{L} &\equiv& \sum_{i=1}^{2N} f_{i , L}^{0} . \label{eq:NLunp}
\end{eqnarray}

From such a quantity, we can easily extract the number of $0$-spin domains with
size $L$ of a particular configuration.  We have thus that the number
of domains of $0$-spins with size $L$ is given by
\begin{eqnarray}
  N^0_{d_{L}} &\equiv& \frac{N^0_{L}}{L} = \sum_{i=1}^{2N} \frac{f_{i , L}^{0}}{L} . \label{eq:NdLunp}
\end{eqnarray}

In addition, if we want to count all the domains of $0$-spins,
irrespective of their size, we just have to sum $N_{d_L}^0$ over all
possible sizes $L$,
\begin{eqnarray}
  N_{d 0} &\equiv& \sum_{L=1}^{2N} N^0_{d_L} = \sum_{L=1}^{2N} \sum_{i=1}^{2N} \frac{f_{i , L}^{0}}{L} . \label{eq:Ndunp}
\end{eqnarray}

Instead of defining the operator counting the total number of domains
of $0$-spins as was done in Eq. (\ref{eq:Ndunp}), we can use an
equivalent and simpler expression for such an operator. It reads 
\begin{eqnarray}
   N_{d 0} &\equiv& \left \{ \begin{array}{r} \frac{N_{\pm 0}}{2} \quad \textrm{if $L \neq 2N$,}
   \\ 1 \hspace{0.6cm} \textrm{if $L = 2N$,} \end{array} \right . 
\end{eqnarray}
where $N_{\pm 0}$ stands for the number of links between polarized and unpolarized
spins, while the term $\delta_{L_{0}, 2 N}$ accounts for the situation
in which all the spins in the chain are unpolarized, in which case
there are no links between polarized and unpolarized spins, but there
is one domain of $0$-spins occupying the entire chain. This operator
can be written explicitly as
\begin{eqnarray}
  N_{d 0} &\equiv& \frac{1}{2} \sum_{i=1}^{2N} \Big[ \sigma_i^2 \big( 1 - \sigma_{i+1}^2 \big) + \big(1 
  - \sigma_i^2 \big) \sigma_{i+1}^2 \Big] \nonumber \\ &&+ \prod_{i=1}^{2 N} \big( 1 - \sigma_{i}^2 \big) 
  . \label{eq:Ndsimp2unp} 
\end{eqnarray} 

Such a definition is introduced because the operator $N_{\pm 0}
/ 2$ is not equivalent to the operator counting the number of domains
in a one-dimensional spin chain. This operator, in fact, counts the
number of links between polarized and unpolarized spins ($+0$, $-0$,
$0+$ and $0-$) divided by two. Whenever the spin configuration is such
that there are links between polarized and unpolarized spins, this
operator is equivalent to the operator giving the number of
domains. However, when there are no links between polarized and
unpolarized spins, this operator always yields $0$, not being able to
distinguish between the cases where all the spins are polarized (and
thus the number of unpolarized spin domains is $N_{d 0} = 0$) and the
case where all the spins are unpolarized (and thus the number of
unpolarized spin domains is $N_{d 0} = 1$). In order to account for
these cases, the term $\prod_{i=1}^{2 N} \big( 1 - \sigma_{i}^2 \big)$
is added to the definition of $N_{d0}$, giving $1$ when the whole spin
chain is unpolarized.

We can write analogous equations to
Eqs. (\ref{eq:NLunp})-(\ref{eq:Ndsimp2unp}) for the case of polarized
spins.  The operator counting the number of polarized spins in domains
of $\pm$-spins, $N^{\pm}_{L}$, reads
\begin{eqnarray}
  N^{\pm}_{L} &\equiv& \sum_{i=1}^{2N} f_{i , L}^{\pm} , \label{eq:NLpol}
\end{eqnarray}
while the operator counting the number of domains (of $\pm$-spins)
with size $L$, reads
\begin{eqnarray}
  N^{\pm}_{d_{L}} &\equiv& \frac{N^{\pm}_{L}}{L} = \sum_{i=1}^{2N} \frac{f_{i , L}^{\pm}}{L} . \label{eq:NdLpol}
\end{eqnarray}
The total number of $\pm$-spins domains, irrespectively of their size,
$N_{d \pm}$, reads
\begin{eqnarray}
  N_{d \pm} &\equiv& \sum_{L=1}^{2N} N^{\pm}_{{d}_L} = \sum_{L=1}^{2N} \sum_{i=1}^{2N} \frac{f_{i , L}^{\pm}}{L} , 
  \label{eq:Ndpol}
\end{eqnarray}
which, in analogy with what was done for $N_{d 0}$, can be rewritten,
reading
\begin{eqnarray}
  N_{d \pm} &\equiv& \frac{1}{2} \sum_{i=1}^{2N} \Big[ \sigma_i^2 + \sigma_{i+1}^2 - \sigma_i^2 \sigma_{i+1}^2 
  + \sigma_i \sigma_{i+1} \Big] + \prod_{i=1}^{2 N} \sigma_{i}^2 . 
  \nonumber \\ \label{eq:Ndsimp2pol} 
\end{eqnarray}
Note that the term $\prod_{i=1}^{2 N} \sigma_{i}^2$, evaluates to $1$
when all the spins are polarized (forming a polarized spin domain with
a length $L = 2 N$) and to $0$ in all other cases.

We can now obtain the DSD of $0$-spins by computing
the thermal average of the ratio $N_{d_L}^{0} / N_{d 0}$ [see
Eq. (\ref{eq:DSDdefunp})]. The computation of thermal averages of
ratios can be performed using the following mathematical trick
\begin{eqnarray}
  \Bigg\langle \frac{N^{0}_{d_L}}{N_{d 0}} \Bigg\rangle &=& \bigg\langle \int_0^{\infty} N^{0}_{d_L} e^{-u N_{d 0}} 
  \textrm{d} u \bigg\rangle \nonumber \\ 
  &=& \int_0^{\infty} \Big\langle N^{0}_{d_L} e^{-u N_{d 0}} \Big\rangle \textrm{d} u , \label{eq:astuceunp}
\end{eqnarray}
where in the last equality we assumed that we can interchange the integration and
averaging procedures, regardless of the size of the system.

In the above thermal average, given by the sum over all
configurations, we need to exclude the two configurations with all
spins polarized, since $N^{0}_{d_L}=0$ and $N_{d 0}=0$ in such case,
yielding indeterminate terms to the sum. This is equivalent to the
computation of the conditioned probability of having a domain of
unpolarized spins with a particular size $L$, given that there are
domains of $0$-spins in the one-dimensional chain. Excluding these
terms changes the partition function, $Z_{2 N}$, from $Z_{2 N} =
\lambda_{+}^{2 N} + \lambda_{0}^{2 N} + \lambda_{-}^{2 N}$ to $Z'_{2
  N} = Z_{2 N} - 2 e^{-2N\beta \delta} = \lambda_{+}^{2 N} -
\lambda_{0}^{2 N} + \lambda_{-}^{2 N}$. In addition, note that the
sums over all the configurations must also exclude the terms
associated with this configuration.

We can thus rewrite Eq. (\ref{eq:astuceunp}) as
\begin{eqnarray}
  \Bigg\langle \frac{N^0_{d_L}}{N_{d 0}} \Bigg\rangle &=& \int_0^{\infty} \sum_{\{ \sigma \}'} \frac{ 
    N^0_{d_L} e^{-\beta E(\{ \sigma \}) - u N_{d 0}}}{Z'_{2 N}} \textrm{d} u , \label{eq:astuceunp2}
\end{eqnarray}
where $\{\sigma\}'$ indicates that the sum is performed over all the
configurations except the two configurations with all spins
polarized. Note, however that in Eq. (\ref{eq:astuceunp2}), summing
over $\{\sigma\}'$ or over all the configurations, $\{\sigma\}$,
yields the same result, because the two configurations with all spins
polarized contribute with $N^0_{d_L} = 0$ to the sum.

The version of Eq. (\ref{eq:astuceunp2}) for domains of polarized
spins is obtained by substituting in Eq. (\ref{eq:astuceunp2})
$N_{d_L}^0$ and $N_{d 0}$ by respectively, $N_{d_L}^{\pm}$ and $N_{d
  \pm}$, while $Z'_{2 N} = Z_{2 N} - 1 = \lambda_{+}^{2 N} +
\lambda_{0}^{2 N} + \lambda_{-}^{2 N} - 1$, since in this case the
configuration yielding $N_{d \pm} = 0$ is that with all the spins
unpolarized. Explicitly, it reads
\begin{eqnarray}
  \Bigg\langle \frac{N^{\pm}_{d_L}}{N_{d \pm}} \Bigg\rangle &=& \int_0^{\infty} \sum_{\{ \sigma \}'} \frac{ 
    N^{\pm}_{d_L} e^{-\beta E(\{ \sigma \}) - u N_{d \pm}}}{Z'_{2 N}} \textrm{d} u , \label{eq:astucepol2}
\end{eqnarray}

\subsubsection{The exact expression of the DSD}

In order to obtain the exact expression of the DSD of $0$-spins, we need to compute
the integrand of Eq.  (\ref{eq:astuceunp2}). The sum over all
configurations in Eq. (\ref{eq:astuceunp2}) can still be performed
using the transfer matrix formalism (see Appendix
\ref{appA}). However, here we have to use a modified transfer matrix, 
which absorbs the exponential of the number of domains $N_{d 0}$ appearing
in Eq. (\ref{eq:astuceunp2}) in the definition given by Eq. (\ref{eq1}). 
It reads
\begin{eqnarray}
  \widetilde{T} &=& \left( \begin{array}{ccc} 0 & e^{-\beta \widetilde{\gamma}} & e^{-\beta \delta}  \\ 
      e^{-\beta \widetilde{\gamma}} & 1 & e^{-\beta \widetilde{\gamma}} \\
      e^{-\beta \delta} & e^{-\beta \widetilde{\gamma}} & 0 \end{array} \right) , \label{eq:Tmatrix-mod}
\end{eqnarray}
where we have rescaled the exchange parameter $\gamma$ to
$\widetilde{\gamma} = \gamma + u / (2 \beta)$.  Both the eigenvalues,
$\widetilde{\lambda}_+$, $\widetilde{\lambda}_0$,
$\widetilde{\lambda}_-$ and the eigenvectors, $| \widetilde{\lambda}_0
\rangle$, $| \widetilde{\lambda}_0 \rangle$, $| \widetilde{\lambda}_0
\rangle$, of this $\widetilde{T}$-matrix, have exactly the same form
of those obtained for the $T$-matrix, with $\gamma$ substituted by
$\widetilde{\gamma}$. However, both the eigenvalues and the eigenvectors
now depend on the integration variable $u$, through the rescaled exchange
parameter $\widetilde{\gamma}$.

As mentioned above, this new $\widetilde{T}$-matrix originates from
the definition of $N_{d 0}$ [see Eq.  (\ref{eq:Ndsimp2unp})]. Note
however, that there is a subtlety in the definition of the
$\widetilde{T}$-matrix in Eq. (\ref{eq:Tmatrix-mod}). In fact, this
transfer matrix absorbs not $e^{- u N_{d 0}}$, but instead $e^{- u
  N_{\pm 0}/2}$ into itself. The factor $e^{- u \delta_{L_0,2N}}$ that
also enters the definition of $N_{d 0}$ [see
Eq. (\ref{eq:Ndsimp2unp})], is not absorbed into
$\widetilde{T}$-matrix, because this term involves all the spins of
the chain, which cannot be properly represented using a nearest
neighbor transfer matrix formalism. As a consequence, we must keep in
mind that the results of the sum over all configurations in
Eq. (\ref{eq:astuceunp2}), will need to include an additional factor
of $e^{-u}$ in the cases where all the spins are unpolarized, i.e. when 
$L_0=2 N$.

For domains of polarized spins, the sum over all configurations in
Eq. (\ref{eq:astucepol2}) is still computed using a modified transfer
matrix. This $\widetilde{T}$-matrix is the same as that of
Eq. (\ref{eq:Tmatrix-mod}), defined for the case of unpolarized
domains. However, if we remember the definition of the operator
counting the number of polarized spins domains [see Eq.
(\ref{eq:Ndsimp2pol})], $N_{d \pm} = N_{\pm 0} / 2 +
\delta_{L_{\pm},2N}$, we readily conclude that now, our results will
need to include an additional factor of $e^{-u}$ in the cases where
all the spins are polarized, i.e. $L_{\pm}=2 N$, and not when $L_0=2 N$.

Given this, we can rewrite the integrand in Eq. (\ref{eq:astuceunp2})
for the DSD of unpolarized spins, using the transfer matrix
formalism as,
\begin{eqnarray}
  I_{0}(L) &=& \frac{1}{Z'_{2N}} \frac{\widetilde{Z}_{2 N}}{L} \sum_{i=1}^{2N} \big\langle f_{i , L}^{0} 
  \big\rangle_{\widetilde{T}} e^{-u \delta_{L,2 N}} , \label{eq:int1} 
\end{eqnarray}
where $\widetilde{Z}_{2 N}$ is the partition function associated with
the $\widetilde{T}$-matrix.  In what concerns the computation of the
DSD of $\pm$-spin domains, note that the integrand in
Eq. (\ref{eq:astucepol2}), $I_{\pm}(L)$, is of the same form as
$I_{0}(L)$ on Eq. (\ref{eq:int1}), but with $f_{i, L}^{0}$ substituted
by $f_{i, L}^{\pm}$,
\begin{eqnarray}
  I_{\pm}(L) &=& \frac{1}{Z'_{2N}} \frac{\widetilde{Z}_{2 N}}{L} \sum_{i=1}^{2N} \big\langle f_{i , L}^{\pm} 
  \big\rangle_{\widetilde{T}} e^{-u \delta_{L,2 N}} , \label{eq:intPol1} 
\end{eqnarray}
where we should recall that the $Z'_{2 N}$ in Eq. (\ref{eq:intPol1})
is different from that appearing in Eq. (\ref{eq:int1}). In addition,
note that the exponential term in Eq. (\ref{eq:intPol1}) refers to the
configuration where all the spins of the $1$D chain are polarized,
while such term in Eq. (\ref{eq:int1}) refers to the configuration where
all the spins are unpolarized.

If we now define $\Theta(L) = \xi_i \xi_{i+1} \ldots \xi_{i+L-1}$
where $\xi_i \equiv 1 - \sigma_i^2$, we can, using
Eqs. (\ref{eq:f0i1})-(\ref{eq:f0i3}), write Eq. (\ref{eq:int1}) as
\begin{subequations} \label{eq:int2}
  \begin{eqnarray}
    I_{0}(L\leq2N-2) &=& 2 N \frac{\widetilde{Z}_{2 N}}{Z'_{2 N}} \Big[ \langle \Theta(L+2) \rangle_{\widetilde{T}} 
    \nonumber \\ &-& 2 \langle \Theta(L+1) \rangle_{\widetilde{T}} + \langle \Theta(L) \rangle_{\widetilde{T}} 
    \Big] , \\
    I_{0}(L=2N-1) &=& 2 N \frac{\widetilde{Z}_{2 N}}{Z'_{2 N}} \Big[ \langle \Theta(2N) \rangle_{\widetilde{T}} 
    \nonumber \\ &-& \langle \Theta(2N-1) \rangle_{\widetilde{T}} \Big] , \\
    I_{0}(L=2N) &=& \frac{\widetilde{Z}_{2 N}}{Z'_{2 N}} \langle \Theta(2 N) \rangle_{\widetilde{T}} e^{- u} .
    \label{eq:int2-3}
  \end{eqnarray}
\end{subequations}

In the case of the domains of polarized spins, in accordance with
Eqs. (\ref{eq:fpmi1})-(\ref{eq:fpmi3}), the integrands are obtained
from Eqs. (\ref{eq:int2}), just by substituting $\Theta(L)$ by $\Gamma
(L) = \sigma_i^2 \sigma_{i+1}^2 \ldots \sigma_{i+L-1}^2$,
\begin{subequations} \label{eq:intPol2}
  \begin{eqnarray}
    I_{\pm}(L\leq2N-2) &=& 2 N \frac{\widetilde{Z}_{2 N}}{Z'_{2 N}} \Big[ \langle \Gamma(L+2) \rangle_{\widetilde{T}} 
    \nonumber \\ &-& 2 \langle \Gamma(L+1) \rangle_{\widetilde{T}} + \langle \Gamma(L) \rangle_{\widetilde{T}} 
    \Big] , \\
    I_{\pm}(L=2N-1) &=& 2 N \frac{\widetilde{Z}_{2 N}}{Z'_{2 N}} \Big[ \langle \Gamma(2N) \rangle_{\widetilde{T}} 
    \nonumber \\ &-& \langle \Gamma(2N-1) \rangle_{\widetilde{T}} \Big] , \\
    I_{\pm}(L=2N) &=& \frac{\widetilde{Z}_{2 N}}{Z'_{2 N}} \langle \Gamma(2 N) \rangle_{\widetilde{T}} e^{- u} .
    \label{eq:intPol2-3}
  \end{eqnarray}
\end{subequations}

Computing the correlation functions $\langle \Theta(L)
\rangle_{\widetilde{T}}$, using the transfer matrix formalism involves
the computation of the following trace
\begin{eqnarray}
  \langle \Theta(L) \rangle_{\widetilde{T}} &=& \frac{1}{\widetilde{Z}_{2 N}} \textrm{Tr} \Big[ \widetilde{T}^{2 N-L} 
  (\xi \widetilde{T})^{L} \Big] , \label{eq:ExpTheta}
\end{eqnarray}
which yields the result $\langle \Theta(L) \rangle_{\widetilde{T}} =
\widetilde{F}(1)^{L-1} \widetilde{F}(2 N-L+1) / \big(\widetilde{Z}_{2
  N} \widetilde{F}(0)^L \big)$, where $\widetilde{F}(r) =
\widetilde{\alpha}_{+} \widetilde{\beta}_{-}
\widetilde{\lambda}_{-}^{r} - \widetilde{\alpha}_{-}
\widetilde{\beta}_{+} \widetilde{\lambda}_{+}^{r}$. The
$\widetilde{\alpha}_{\pm}$ and $\widetilde{\beta}_{\pm}$ are the
entries of the eigenvectors of $\widetilde{T}$, $| \lambda_{\pm}
\rangle$ [see Eq. (\ref{eqA5})]. Noting that $\widetilde{F}(1) =
\widetilde{F}(0)$ and using Eq. (\ref{eqA5}), we can finally write
$\langle \Theta(L) \rangle_{\widetilde{T}}$ as
\begin{eqnarray}
  \langle \Theta(L) \rangle_{\widetilde{T}} &=& \frac{1}{\widetilde{Z}_{2 N}} \frac{ (\widetilde{\lambda}_{+} - 1) 
    \widetilde{\lambda}_{-}^{p} - (\widetilde{\lambda}_{-} - 1) \widetilde{\lambda}_{+}^{p} }
  {\sqrt{(e^{-\beta \delta} - 1)^2 + 8 e^{-2 \beta \widetilde{\gamma}}}} , \label{eq:SolTheta2}
\end{eqnarray}
where $p = 2 N - (L-1)$.

In the computation of the DSD of the polarized spins, the
$\langle \Gamma(L) \rangle_{\widetilde{T}}$ appearing in
Eqs. (\ref{eq:intPol2}) can be analogously computed and one obtains
\begin{eqnarray}
  \langle \Gamma(L) \rangle_{\widetilde{T}} &=& \frac{e^{-\beta \delta (L-1)}}{\widetilde{Z}_{2 N}}
  \frac{ (\widetilde{\lambda}_{+} - 1) \widetilde{\lambda}_{+}^{p} - (\widetilde{\lambda}_{-} - 1) 
    \widetilde{\lambda}_{-}^{p} }{\sqrt{(e^{-\beta \delta} - 1)^2 + 8 e^{-2 \beta \widetilde{\gamma}}}} \nonumber \\
  &+& \frac{\lambda_{0}^{2 N}}{\widetilde{Z}_{2 N}} ,  \label{eq:SolGamma2}
\end{eqnarray}
where, again, $p = 2 N - (L-1)$.

The integrands in Eqs. (\ref{eq:int2}) can be rewritten as
\begin{subequations} \label{eq:I0s}
  \begin{eqnarray}
    I_{0}(L\leq2N-2) &=& 2 N \frac{1}{Z'_{2 N}} \Big( W_- - W_+ \Big) , \\
    I_{0}(L=2N-1) &=& 2 N \frac{1}{Z'_{2 N}} \Big( Y_- - Y_+ \Big) , \\
    I_{0}(L=2N) &=& \frac{1}{Z'_{2 N}} e^{- u} ,
  \end{eqnarray} \label{eq:int3}
\end{subequations}
where $W_{\pm} = \widetilde{\lambda}_{\mp}^{p-2}
(\widetilde{\lambda}_{\pm} - 1) (\widetilde{\lambda}_{\mp} - 1)^2 /
(\widetilde{\lambda}_+ - \widetilde{\lambda}_-)$, while $Y_{\pm} =
\widetilde{\lambda}_{\mp}^{p-1} (\widetilde{\lambda}_{\pm} - 1)
(\widetilde{\lambda}_{\mp} - 1) / (\widetilde{\lambda}_+ -
\widetilde{\lambda}_-)$.

For the polarized spins domains case, the integrands in
Eqs. (\ref{eq:intPol2}) can be rewritten as
\begin{subequations} \label{eq:IPMs}
  \begin{eqnarray}
    I_{\pm}(L\leq2N-2) &=& 2 N \frac{1}{Z'_{2 N}} \Big( \mathcal{W}_- - \mathcal{W}_+ \Big) , \\
    I_{\pm}(L=2N-1) &=& 2 N \frac{1}{Z'_{2 N}} \Big( \mathcal{Y}_- - \mathcal{Y}_+ \Big) , \\
    I_{\pm}(L=2N) &=& \frac{2 \lambda_{0}^{2 N}}{Z'_{2 N}} e^{- u} ,
  \end{eqnarray} \label{eq:intPol3}
\end{subequations}
where $\mathcal{W}_{\pm} = e^{-\beta \delta (L-1)}
\widetilde{\lambda}_{\pm}^{p-2} (\widetilde{\lambda}_{\pm} - 1)
(\widetilde{\lambda}_{\pm} - e^{-\beta \delta})^2 /
(\widetilde{\lambda}_+ - \widetilde{\lambda}_-)$, while
$\mathcal{Y}_{\pm} = e^{-\beta \delta (L-1)}
\widetilde{\lambda}_{\pm}^{p-1} (\widetilde{\lambda}_{\pm} - 1)
(\widetilde{\lambda}_{\pm} - e^{-\beta \delta}) /
(\widetilde{\lambda}_+ - \widetilde{\lambda}_-)$.

Performing the integral over $u$ in the expressions for $I_{0}(L)$ as
given in Eqs.(\ref{eq:I0s}), leaves us with the following expressions
for the DSD of unpolarized spins
\begin{widetext}
  \begin{subequations} \label{eq:DSDunpExact}
    \begin{eqnarray}
      \Bigg\langle \frac{N^{0}_{d_L}}{N_{d 0}} \Bigg\rangle_{L\leq2N-2} &=& \frac{2 N}{Z'_{2N}} \frac{1}{2^{m+2}} \Bigg[ 
      \frac{1}{m+1} \bigg\{ c \Big( G_-^{m+1} + G_+^{m+1} \Big) - \sqrt{c^2 + d} 
      \Big( G_-^{m+1} - G_+^{m+1} \Big) - 2 c \Big( 2 (c+1) \Big)^{m+1} \bigg\} \nonumber \\
      &-& \frac{1}{(m+1) (m+2)} \bigg\{ \Big( G_-^{m+2} + G_+^{m+2} \Big) - 2^{m+2} \Big( 1 + (c+1)^{m+2} \Big) 
      \bigg\} \Bigg], \label{eq:DSDunpExact1} \\
      \Bigg\langle \frac{N^{0}_{d_L}}{N_{d 0}} \Bigg\rangle_{L=2N-1} &=& \frac{2 N}{Z'_{2N}} \frac{1}{8} \bigg[ \Big(
      G_-^2 + G_+^2 \Big) - 4 \Big( 1 + (c+1)^2 \Big) \bigg], \label{eq:DSDunpExact2} \\
      \Bigg\langle \frac{N^{0}_{d_L}}{N_{d 0}} \Bigg\rangle_{L=2N} &=& \frac{1}{Z'_{2N}} , \label{eq:DSDunpExact3}
    \end{eqnarray}
  \end{subequations}
\end{widetext}
with $m = p - 2$, $G_{\pm} = c + 2 \pm \sqrt{c^2 + d}$, $c = e^{-\beta
  \delta} - 1$ and $d = 8 e^{- \beta \gamma}$. Recall that, as the
above equations refer to the computation of the DSD of
$0$-spins, in Eqs. (\ref{eq:DSDunpExact}) we have that $Z'_{2 N} =
\lambda_{+}^{2 N} - \lambda_{0}^{2 N} + \lambda_{-}^{2 N}$.

The DSD of polarized spins, is analogously given from
Eqs.(\ref{eq:IPMs}), by
\begin{widetext}
  \begin{subequations} \label{eq:DSDpolExact}
    \begin{eqnarray}
      \Bigg\langle \frac{N^{\pm}_{d_L}}{N_{d \pm}} \Bigg\rangle_{L\leq2N-2} &=& \frac{2 N}{Z'_{2N}} 
      \frac{e^{-\beta \delta (L-1)}}{2^{m+2}} \Bigg[ 
      \frac{1}{m+1} \bigg\{ \bar{c} \Big( \mathcal{G}_-^{m+1} + \mathcal{G}_+^{m+1} \Big) - \sqrt{\bar{c}^2 + d} 
      \Big( \mathcal{G}_-^{m+1} - \mathcal{G}_+^{m+1} \Big) - 2^{2 m + 2} \bar{c} \bigg\} \nonumber \\
      &-& \frac{1}{(m+1) (m+2)} \bigg\{ \Big( \mathcal{G}_-^{m+2} + \mathcal{G}_+^{m+2} \Big) - 2^{m+2} 
      \Big( 1 + (1 - \bar{c})^{m+2} \Big) \bigg\} \Bigg], \label{eq:DSDpolExact1} \\
      \Bigg\langle \frac{N^{\pm}_{d_L}}{N_{d \pm}} \Bigg\rangle_{L=2N-1} &=& \frac{2 N}{Z'_{2N}} 
      \frac{e^{-\beta \delta (2 N-2)}}{8} \bigg[ \Big(
      G_-^2 + G_+^2 \Big) - 4 \Big( 1 + (1 - \bar{c})^2 \Big) \bigg], \label{eq:DSDpolExact2} \\
      \Bigg\langle \frac{N^{\pm}_{d_L}}{N_{d \pm}} \Bigg\rangle_{L=2N} &=& \frac{2 \lambda_{0}^{2 N}}{Z'_{2N}} , 
      \label{eq:DSDpolExact3}
    \end{eqnarray}
  \end{subequations}
\end{widetext}
with $\bar{c} = 1 - e^{-\beta \delta}$ and $\mathcal{G}_{\pm} = 2 -
\bar{c} \pm \sqrt{\bar{c}^2 + d}$.  As Eqs. (\ref{eq:DSDpolExact})
refer to the DSD of $\pm$-spins, the $Z'_{2 N}$ appearing
in Eqs. (\ref{eq:DSDpolExact}) reads $Z'_{2 N} = \lambda_{+}^{2 N} +
\lambda_{0}^{2 N} + \lambda_{-}^{2 N} - 1$.

\subsubsection{The thermodynamic limit of the DSD}

In the thermodynamic limit ($2 N \to \infty$), when the domain size is
much smaller than the size of the system ($L \ll 2 N$), we can perform
the limit $m \to + \infty$ in Eq. (\ref{eq:DSDunpExact1}).  Note that
as $G_+ > G_-$, $c + 1 > 1$ and $G_+ > 2 (c + 1)$, then we have that
the DSD of unpolarized spins, in the thermodynamic limit,
is given by Eq. (\ref{eq:DSDunpTh}). Analogously, the DSD 
of polarized spins, also in the thermodynamic limit, is given by
Eq. (\ref{eq:DSDpolTh}).

\begin{figure}[htp!]
  \centering
  \includegraphics[width=0.98\columnwidth]{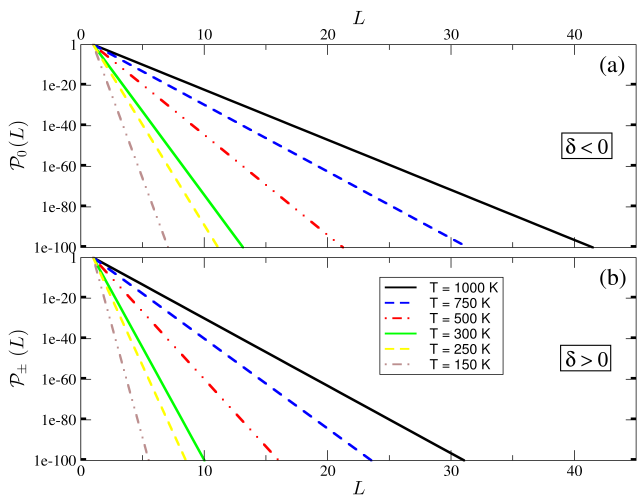}
  \caption{(Color online) Plots of the DSD for two sets of values of
    the exchange parameters $\gamma$ and $\delta$.  In both panels we
    present the DSD (with a logarithmic scale in the $y$-axis) for six
    different temperatures: full black lines stand for $T = 1000 K$;
    dashed blue lines stand for $T = 750 K$; dashed-dotted red lines
    stand for $T = 500 K$; full light green lines stand for $T = 300
    K$; dashed light yellow lines stand for $T = 250 K$; dashed-dotted
    light brown lines stand for $T = 150 K$. (a) DSD of
    unpolarized spins, in an unpassivated edge ($\gamma = 0.03$ and
    $\delta = -0.49$).  (b) DSD of polarized spins, in a
    hydrogen-passivated edge ($\gamma = 0.52$ and $\delta = 0.66$).}
  \label{fig:DSD}
\end{figure}

In Fig. \ref{fig:DSD}, we have plotted the DSD of
unpolarized spins at a non-passivated edge, as well as the DSD of
polarized spins at a hydrogen-passivated one, for two sets
of particular values of the exchange parameters. In these plots we
have used, the values for the exchange parameters $\gamma$ and
$\delta$ computed in Appendix \ref{sec:ExchangeParamsResults}, i.e.
$\gamma = 0.03$ and $\delta = -0.49$ for the unpassivated edge and
$\gamma = 0.52$ and $\delta = 0.66$ for the hydrogen-passivated
edge. We readily see that both DSDs strongly decrease with increasing
sizes of the domains. Note the dependence of such decrease rate with the
temperature: the larger the temperature is, the smaller the decrease rate
is.

A distribution $P(x)$ is said to have fat tails if it displays a slower
decrease than the normal distribution, (or, alternatively, if it
decreases with a power of $x$) when $x \to \infty$. As a consequence,
the moments of a fat-tailed distribution diverge above a given order,
characteristic of that distribution, and thus its characteristic
function is not analytical at the origin.

Let us now consider the question of whether the DSDs computed above
display fat tails in the thermodynamic limit (let us represent the DSD
generically as $\mathcal{P}(L)$).  Its characteristic function is
given by the discrete Fourier transform
\begin{eqnarray}
  \hat{\mathcal{P}}(w) &=& \sum_{n=1}^{+\infty} e^{i w n} \mathcal{P}(L) .
\end{eqnarray}
The characteristic functions of $\mathcal{P}_{0}(L)$ and of
$\mathcal{P}_{\pm}(L)$ are geometric series, and hence easily
computable, after which we obtain Eqs. (\ref{eq:CharactFuncUnp}) and
(\ref{eq:CharactFuncPol}).
From Eq. (\ref{eq:CharactFuncUnp}), or its geometric series form, we
conclude that all the derivatives of $\hat{\mathcal{P}}_0(w)$ exist at
$w=0$, if $\lambda_+ > 1$. If the unpolarized spins are the minority
spins in the chain, $\delta < 0$ and we have $\lambda_+ \geq 1 +
\sqrt{2} > 1$ for every $\beta \geq 0$, thus $\mathcal{P}_{0}(L)$ has
no fat-tails.
Similarly, from Eq. (\ref{eq:CharactFuncPol}), when the polarized
spins are the minority spins in the spin chain, $\delta > 0$ and we
conclude that all the derivatives of $\hat{\mathcal{P}}_{\pm}(w)$
exist at $w=0$, if $\lambda_{+} e^{\beta \delta} > 1$. This is
verified for every $\beta \geq 0$, since $\lambda_+ e^{\beta \delta}
\geq 1 + \sqrt{2} > 1$. Again, we conclude that $\mathcal{P}_{\pm}(L)$
has no fat-tails.

The first moment of $\mathcal{P}_0(L)$ gives us the mean size of the
domains of unpolarized spins in the thermodynamic limit,
$\bar{L}_{0}$, which reads
\begin{eqnarray}
  \bar{L}_0 &=& - i \frac{\textrm{d}}{\textrm{d} w} \hat{\mathcal{P}}_0(w)\bigg|_{w=0} \,.
  \label{eq:MomDist}
\end{eqnarray}
In the same way, we can compute the mean size of the domains of
polarized spins (in the thermodynamic limit), $\bar{L}_{\pm}$. Both
these two quantities are written, respectively, in
Eqs. (\ref{eq:LmedExactUnp}) and (\ref{eq:LmedExactPol}). These two
quantities are plotted in Fig. \ref{fig:Lmed0}.


%

\end{document}